\documentclass[sigconf]{acmart}
\AtBeginDocument{%
  }

\acmConference[Pre-print]{}{Jan 1}{2026}
\acmBooktitle{}
\renewcommand\footnotetextcopyrightpermission[1]{}
\usepackage{xspace}
\usepackage{amsmath}
\usepackage{xcolor}
\usepackage{booktabs}
\usepackage{subcaption}
\usepackage{listings}
\usepackage{tabularx}
\usepackage{array}
\usepackage{ragged2e}
\usepackage{multirow}
\usepackage[most]{tcolorbox}
\usepackage{comment}
\usepackage[inline]{enumitem}
\usepackage{pifont}
\usepackage{hyperref}
\newcommand{\sysname}{MAESTRO\xspace}

\newcommand{\para}[1]{\vspace{1mm}\noindent\textbf{#1}}

\lstdefinelanguage{json}{
    basicstyle=\ttfamily\small,
    numbers=left,
    numberstyle=\tiny,
    stepnumber=1,
    numbersep=6pt,
    showstringspaces=false,
    breaklines=true,
    frame=single,
    backgroundcolor=\color{gray!5},
    stringstyle=\color{blue},
    keywordstyle=\color{black},
    commentstyle=\color{gray},
}

\newcolumntype{C}[1]{>{\centering\arraybackslash}p{#1}}
\newcolumntype{L}[1]{>{\RaggedRight\arraybackslash}p{#1}}

\newcommand{\fakepara}[1]{\vspace{1mm}\noindent\textbf{#1}\xspace}

\definecolor{rqframe}{HTML}{01b050} 
\definecolor{rqback}{HTML}{f8fcf6}  
\newcounter{rqfinding}
\newcommand{\rqbox}[1]{%
  \refstepcounter{rqfinding}%
  \tcbox[
    capture=minipage,
    width=\linewidth,
    colback=rqback,
    colframe=rqframe,
    boxrule=1.1pt,
    arc=1mm,
    left=6pt,right=6pt,top=5pt,bottom=5pt,
    boxsep=0pt,
    enhanced,
    breakable
  ]{%
    {\itshape Finding~\therqfinding: }#1%
  }%
}

\newif\ifsubmission
\submissionfalse

\ifsubmission
\newcommand{\mcnote}[1]{}
\newcommand{\sfnote}[1]{}
\newcommand{\tmnote}[1]{}
\newcommand{\yxcnote}[1]{}
\newcommand{\vanote}[1]{}
\newcommand{\acnote}[1]{}
\newcommand{\afnote}[1]{}
\else
\usepackage{todonotes}
\newcommand{\mcnote}[1]{\todo[color=purple!40,inline]{MC: #1}}
\newcommand{\sfnote}[1]{\todo[color=violet!20,inline]{SF: #1}}
\newcommand{\tmnote}[1]{\todo[color=orange!40,inline]{Tie: #1}}
\newcommand{\vanote}[1]{\todo[color=magenta!40,inline]{VA: #1}}
\newcommand{\yxcnote}[1]{\todo[color=olive!20,inline]{Yixi: #1}}

\newcommand{\acnote}[1]{\todo[color=teal!20,inline]{AC: #1}}
\newcommand{\afnote}[1]{\todo[color=brown!20,inline]{Amandio: #1}}

\fi

\def\Snospace~{\S{}}

\definecolor{figurecolor}{RGB}{22,90,220}
\definecolor{citecolor}{RGB}{198,81,19}
\hypersetup{ colorlinks=true, urlcolor=black, linkcolor=figurecolor, citecolor=citecolor, pdfstartview=FitH,
        pdftitle={\sysname: Multi-Agent Evaluation Suite for Testing, Reliability, and Observability},
        pdfauthor={}
        }

\begin{document}

\title[MAESTRO: Multi-Agent Evaluation Suite for Testing, Reliability, and Observability]{\sysname: Multi-Agent Evaluation Suite for\\ Testing, Reliability, and Observability}

\author{Tie Ma}
\authornote{Equal contribution.}
\authornote{Work done while Tie Ma was interning at KAUST.}
\affiliation{%
  \institution{Beihang University}
  \country{\relax}
}

\author{Yixi Chen}
\authornotemark[1]
\affiliation{%
  \institution{KAUST}
  \country{\relax}
}

\author{Vaastav Anand}
\affiliation{%
  \institution{MPI-SWS}
  \country{\relax}
}

\author{Alessandro Cornacchia}
\affiliation{%
  \institution{KAUST}
  \country{\relax}
}

\author{Amândio R. Faustino}
\affiliation{%
  \institution{KAUST}
  \country{\relax}
}

\author{Guanheng Liu}
\affiliation{%
  \institution{Beihang University}
  \country{\relax}
}

\author{Shan Zhang}
\affiliation{%
  \institution{Beihang University}
  \country{\relax}
}

\author{Hongbin Luo}
\affiliation{%
  \institution{Beihang University}
  \country{\relax}
}

\author{Suhaib A. Fahmy}
\affiliation{%
  \institution{KAUST}
  \country{\relax}
}

\author{Zafar A. Qazi}
\affiliation{%
  \institution{LUMS \& KAUST}
  \country{\relax}
}

\author{Marco Canini}
\affiliation{%
  \institution{KAUST}
  \country{\relax}
}

\renewcommand{\shortauthors}{T. Ma, Y. Chen, et al.}

\begin{abstract}
Large language model (LLM)-based multi-agent systems (MAS) are rapidly moving from demos to production, yet their dynamic execution makes them stochastic, failure-prone, and difficult to reproduce or debug. Existing benchmarks largely emphasize application-level outcomes (e.g., task success) and provide limited, non-standardized visibility into execution behavior, making controlled, apples-to-apples comparisons across heterogeneous MAS architectures challenging.

We present \sysname, an evaluation suite for the testing, reliability, and observability of LLM-based MAS. \sysname standardizes MAS configuration and execution through a unified interface, supports integrating both native and third-party MAS via a repository of examples and lightweight adapters, and exports framework-agnostic execution traces together with system-level signals (e.g., latency, cost, and failures). We instantiate \sysname with 12 representative MAS spanning popular agentic frameworks and interaction patterns, and conduct controlled experiments across repeated runs, backend models, and tool configurations. Our case studies show that MAS executions can be structurally stable yet temporally variable, leading to substantial run-to-run variance in performance and reliability. We further find that MAS architecture is the dominant driver of resource profiles, reproducibility, and cost--latency--accuracy trade-off, often outweighing changes in backend models or tool settings. Overall, \sysname enables systematic evaluation and provides empirical guidance for designing and optimizing agentic systems.\end{abstract}

\maketitle

\section{Introduction}
\label{sec: intro}

\begin{table*}[t]
    \centering
    \caption{Summary of Systematic Findings across Case Studies (§4)}
    \label{tab:systematic-findings}
    \begin{tabularx}{\textwidth}{l l X} 
        \toprule
        \textbf{Subject} & \textbf{Ref.} & \textbf{Finding} \\ 
        \midrule
        
        \textbf{Resources} & \autoref{sec:mas-overall-res-consumption} & Execution requires minimal resources: sub-GB memory, <20\% of a CPU core, and MB-scale traffic. \\ 
        
        \midrule
        
        \multirow{2}{*}{\textbf{General}} 
        & \autoref{sec:mas-graph-stability} & Interaction structures remain stable while call sequences exhibit temporal instability. \\ \cmidrule(lr){2-3}
        & \autoref{sec:mas-tool-impact} & Tool integration mitigates speculative generation, reducing latency and cost. \\
        
        \midrule
        
        \multirow{2}{*}{\textbf{Backend}} 
        & \autoref{sec:mas-model-impact} & Model scaling yields inconsistent gains; execution dynamics dominate performance. \\ \cmidrule(lr){2-3}
        & \autoref{sec:mas-failure-attribution} & Model-specific failures are significantly amplified by execution dynamics. \\
        
        \midrule
        
        \multirow{4}{*}{\textbf{Architecture}} 
        & \autoref{sec:mas-overall-res-consumption} & MAS architecture significantly dominates resource consumption profiles. \\ \cmidrule(lr){2-3}
        & \autoref{sec:mas-graph-stability} & Architecture governs call graph similarity and determines system reproducibility. \\ \cmidrule(lr){2-3}
        & \autoref{sec:mas-arch-impact} & Generalized architectures incur higher resource overhead without accuracy gains. \\ \cmidrule(lr){2-3}
        & \autoref{sec:mas-tool-impact} & Accuracy gains are architecture-dependent and contingent on low execution overhead. \\
        
        \bottomrule
    \end{tabularx}
\end{table*}

LLM-based multi-agent systems (MAS) enable flexible task solvers that can handle diverse and multimodal workloads~\cite{ijcai2024p890} with minimal modification to the underlying system architecture. 
However, this flexibility also introduces substantial uncertainty in system load and execution behavior. 
Unlike traditional deterministic workflows, LLM-based MAS operate under a dynamic execution model in which decisions are made on the fly during runtime, driven by LLM outputs rather than by statically defined control flow.

Importantly, MAS should not be viewed merely as a collection of lightweight client-side frameworks. 
Instead, they constitute complex systems characterized by dynamic interactions~\cite{yan2025beyond}, emergent behaviors~\cite{ijcai2024p890}, and a broad spectrum of failure modes~\cite{cemri2025multi}.
These characteristics challenge conventional assumptions such as predictability, observability, and performance isolation, making traditional system optimization techniques less effective in this context.
Therefore, a benchmark suite that systematically characterizes MAS execution behavior is essential for both system operators seeking performance optimization and researchers aiming to identify open challenges and opportunities for innovation.

Unfortunately, existing standardized benchmarks for LLM-based MAS remain limited and often lack broad coverage of MAS execution behavior.
Prior work has largely focused on LLM serving and inference efficiency~\cite{wang2025burstgpt,chitty2024llm,liang2023holistic,srivastava2023beyond,hendrycks2020measuring,alpaca_eval}, evaluating server-side model performance rather than the execution behavior of agent systems.
With the emergence of LLM-based MAS, recent benchmarks~\cite{liu2023agentbench,bogavelli2025agentarch,guo2024stabletoolbench,yao2024tau,geng2025realm,yan2025beyond,basu2025nestful,juneja2025magpie,sun2025collab,wang2024battleagentbench} have begun to assess individual agent capabilities (e.g., tool use and communication strategies); however, they largely remain centered on application-level performance (e.g., task success and response quality) and fall short of offering a standardized, comprehensive observability perspective on the system-level impact of MAS execution and corresponding workload management challenges.
This fragmentation makes it difficult to reason about complex runtime behavior and to compare systems consistently across settings.
Consistent with this gap, a recent survey~\cite{pan2025measuring} reports that nearly 75\% of teams operating production MAS evaluate their systems without benchmarks, while 25\% build custom benchmarks, limiting portability and reuse across scenarios.

Based on these observations, we define the following core objectives necessary for a good benchmark for LLM-based MAS:

\para{O1: Architectural heterogeneity.}
The execution stack of LLM-based MAS is highly malleable. 
A single objective can be realized through diverse configurations, including the number of agents, role assignments, interaction topologies (e.g., centralized, hierarchical, or peer-to-peer), and communication protocols.
Furthermore, the design space encompasses choices regarding orchestration frameworks, backend LLMs, budget constraints, tool availability, and memory mechanisms, as well as policies for reflection and termination.

\para{O2: Functional representativeness.}
The rapid proliferation of agentic workflows and real-world deployments has led to a growing diversity of MAS architectures, many of which are optimized for specific task patterns or application domains. 
Recent designs explore increasingly sophisticated coordination and reasoning strategies~\cite{yao2023tree, yao2022react, shinn2023reflexion, xu2023rewoo}.
As a result, no single architecture can be considered representative of the broader MAS design space.

\para{O3: Execution traceability.}
Current commercial agentic systems often expose high-level reasoning traces but offer limited, non-standardized visibility into execution-level details and internal system states.
Furthermore, existing MAS modules lack a unified telemetry standard, often resulting in ``silent'' information consumption where different LLM providers and frameworks fail to expose critical operational data to the user~\cite{schoen2025stress}.

To address this gap, we present \sysname, a comprehensive, open-source evaluation suite for LLM-based MAS. 
\sysname is designed to enable systematic characterization of execution behavior across diverse agent architectures, interaction patterns, and runtime conditions, with the goal of informing principled system optimization.

Our contributions  are threefold:

\para{Rich and extensible benchmarks.}
\sysname incorporates 12 representative MAS examples, each characterized by distinct architectural differences, to serve as a foundation for deriving systematic insights, as shown in~\autoref{tab:systematic-findings}.
Moreover, \sysname is designed for extensibility, allowing the community to integrate and reuse existing MAS implementations within our evaluation framework with minimal effort.

\para{Framework-agnostic system integration.}
\sysname is built upon a collection of widely used, open-source agentic frameworks and examples~\cite{langgraph, autogen, adk2025}, aiming to capture common architectural patterns observed in practice rather than favoring a single workflow design.

\para{Unified execution-level telemetry standards.}
\sysname defines and implements a unified telemetry interface designed to capture comprehensive execution data across diverse modules.
This architecture establishes a common protocol that various MAS components can conform to, ensuring consistent and transparent monitoring throughout the system lifecycle.

\sysname is available at \url{https://github.com/sands-lab/maestro}.
\section{Background}
\subsection{Anatomy of an LLM-based MAS}
\label{subsec: Anatomy of an LLM-based MAS}

LLM-based Multi-Agent Systems (MAS) are collections of LLM agents that
operate in tandem to complete large tasks that are beyond the capabilities of
individual agents~\cite{guo2024large}.
In a typical MAS, multiple specialized agents collaborate together to plan,
coordinate, and execute large tasks with each individual agent focusing
on a specific sub-task.

\fakepara{Building blocks.} An LLM agent is an entity that autonomously executes
multi-step tasks by combining generative foundational models with external tools, memory, and 
reasoning and planning capabilities~\cite{wang2023llmagent}. 
Agents are designed to operate autonomously in highly dynamic environments where adaptability and strategic decision making are essential.
Each agent is comprised of four key parts:
\begin{enumerate*}[label=(\roman*)]
\item inputs that may include user instructions, developer-specified constraints, multimodal observations, retrieved knowledge, and internal state;
\item a generative Large Language Model (LLM) that maps the current state to decisions;
\item an action interface that enables tool interactions such as data retrieval, API calls, code execution;
\item outputs including user-facing responses and structured actions and artifacts along with updated state.    
\end{enumerate*}

\fakepara{Orchestration and deployment.}
Practitioners orchestrate MAS through workflows written in agentic programming frameworks
such as LangGraph~\cite{langgraph}, CrewAI~\cite{crewai}, AutoGen~\cite{autogen}, LlamaIndex~\cite{llamaindex}, and Agno~\cite{agno}.
Despite the popularity of these third-party frameworks according to surveys~\cite{pan2025measuring,moshkovich2025beyond}, 
detailed interviews with practitioners revealed that practitioners
preferred to build agentic applications from scratch in 85\% of the cases~\cite{pan2025measuring}.
These workflows may be static or dynamic depending on the degree of autonomy
allowed by developers in these systems.
Currently, MAS are deployed as single monolithic applications; however, they are increasingly
developed and deployed as distributed applications~\cite{kagent,cornacchia2025dmas}.

\fakepara{Workflow structure.} MAS workflows often follow a hierarchical structure with task
structures as a tree of sub-tasks. Individual sub-tasks follow a mix of sequential
and parallel flows. Workflows may also contain recursive calls for individual agents~\cite{moshkovich2025beyond}.

\fakepara{Failure types.} MAS applications showcase three main failure types --- System Design Issues, Inter-Agent Misalignment, Task Verification~\cite{cemri2025multi}.
System design issues include configuration issues, API and system issues, and resource mismanagement.
Inter-agent misalignment issues result from a breakdown in critical information flow
from inter-agent interaction and coordination during execution.
This includes planning and coordination errors, incorrect output generation, individual
LLM hallucinations, and incorrect information processing.
Task Verification failures arise when verification strategies are inadequate at identifying
issues.

\fakepara{Sources of non-determinism.} Due to the dynamic and heterogeneous nature
of MAS applications, they exhibit non-determinism due to a multitude of reasons.
First, LLMs are stochastic in nature and often produce different outputs for the same input.
Second, external tool executions are not pre-planned or programmed. 
Additionally, tools may produce non-deterministic results.
Third, workflows are dynamic and change at runtime~\cite{yan2025beyond}.
Non-determinism in dynamic workflows may further be exacerbated due to the availability of agents.
Fourth, built-in reliability mechanisms impact the performance and structure of MAS executions.
For example, quality-driven retries change the execution graph.

\fakepara{Reliability as a first-class citizen.} Typical MAS applications
treat reliability as a first-class citizen as part of the design and implementation
of these systems. They do so in multiple ways.
First, most MAS applications rely on Human-in-the-loop evaluation,
with almost half of the applications executing fewer than five steps before seeking
human-in-the-loop evaluation~\cite{pan2025measuring}.
In addition, developers often augment applications with LLM-as-a-judge
to automate quality checks.
MAS applications also automate retries to improve quality if quality checks fail.
Second, practitioners prioritize quality over real-time responsiveness,
with 66\% of respondents to a recent survey allowing response times of more than a minute~\cite{pan2025measuring}.
Third, practitioners prefer static workflows over dynamic workflows to constrain the autonomy of deployed agents~\cite{pan2025measuring}.

\subsection{Limitations of existing benchmarks}
\label{sec:exist-limit}

Evaluating and benchmarking the performance of Large Language Models has been an important aspect in measuring the efficiency and efficacy of LLMs at executing real-world tasks~\cite{liang2023holistic,srivastava2023beyond,hendrycks2020measuring,alpaca_eval}.
With the recent rise of LLM agents and MAS applications, benchmarking the performance of agentic systems has garnered a great deal of interest from the scientific community.

\fakepara{Agent benchmarks.} Typical agent benchmarks evaluate capabilities of individual LLMs as agents~\cite{liu2023agentbench,bogavelli2025agentarch, lai2022fedscale}.
These benchmarks have been further extended to multi-agent settings.
To do so, researchers have developed specialized benchmarks that evaluate a specific property of agentic systems
such as Tool Calling~\cite{guo2024stabletoolbench,yao2024tau}, Task Planning~\cite{geng2025realm}, communication strategies~\cite{yan2025beyond},
sequential flows~\cite{basu2025nestful}, privacy preservation~\cite{juneja2025magpie}, and collaboration efficacy~\cite{sun2025collab,wang2024battleagentbench}. 
Such specialized benchmarks solely focus on one specific property or dimension of MAS applications
and lack the holistic view required to effectively understand the end-to-end emergent behavior and performance of MAS applications.

\fakepara{Bespoke benchmarks.} Due to the lack of standardization and the diversity
of MAS design space, MAS application developers instead opt to create
custom benchmarks specific to their application.
For example, authors of Autogen~\cite{wu2024autogen} created a bespoke benchmark called Autogenbench~\cite{autogenbench} for tasks developed in the Autogen framework.
According to a recent survey, 25\% of teams for production MAS applications construct
custom benchmarks for their applications, 75\% of teams evaluate their agents
without formal benchmarks and instead rely on A/B testing and direct expert/user feedback~\cite{pan2025measuring}.
Although these benchmarks are suitable for a given specific application,
such benchmarks do not capture the diversity of the MAS design space and do not provide
insight in a broad setting.

\fakepara{Observability tools and benchmarks.} Observability tools and observability-based benchmarks such as Opik~\cite{opik2024}, TRAIL~\cite{deshpande2025trail}, TAMAS~\cite{moshkovich2025beyond}
capture spans and traces of MAS executions which developers use to further analyze
traces to triage issues and to understand MAS executions. 
Beyond standard metrics, such as agent call frequency, external API usage, and per-call token costs, there remains a significant gap in deep application-semantic telemetry.
Addressing this requirement involves capturing granular retry logic details (e.g., attempt counts, triggers, and parent span IDs), agent-specific status conditions (e.g., failure categorization and error reasoning), and output quality assessments.
Such telemetry is essential for providing the execution-level transparency needed to diagnose stochastic failures and understand complex multi-agent interactions.

\section{\sysname}
\label{sec:maestro}

We present \sysname, a Multi-Agent Evaluation Suite for Testing, Reliability, and Observability, as a comprehensive framework for evaluating LLM-based MAS.
Building upon goals, we first outline in \autoref{sec: intro}, we detail the architecture and design of the framework (\autoref{sec:sys-arch}), illustrating how standalone MAS implementations are adapted and integrated into our suite.
To demonstrate \sysname's capacity for generating informative telemetry, we present a collection of representative MAS instances (i.e., the concrete evaluation units in a benchmark suite).
These are categorized according to our proposed taxonomy (\autoref{subsec:mas-taxonomy}), while \autoref{subsec:example-studied} details the specific instances used and the formulation of evaluation suites designed to derive our experimental findings.

\subsection{Benchmark design}
\label{sec:sys-arch}

\subsubsection{\sysname architecture}

\begin{figure}[t]
\centering
\includegraphics[width=0.8\columnwidth]{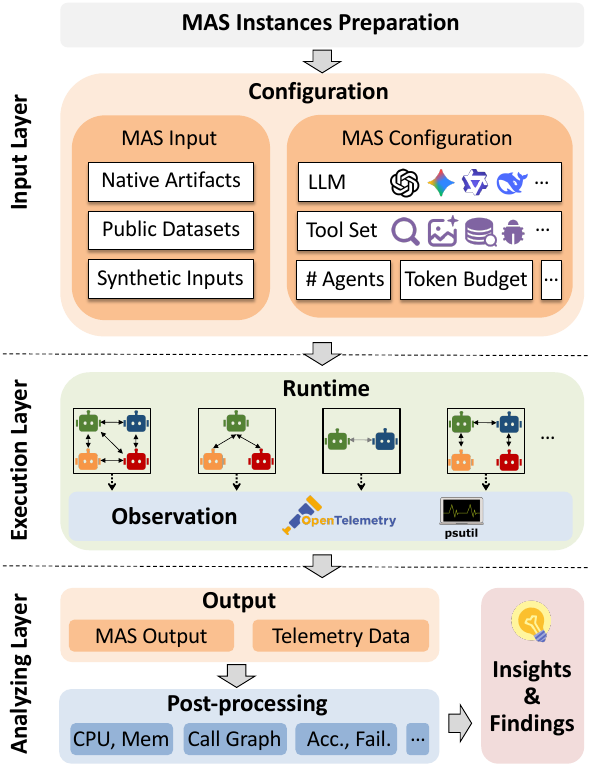}
\caption{\sysname architecture overview.}
\label{fig:arch}
\end{figure}

\autoref{fig:arch} presents an overview of the \sysname architecture.
Conceptually, \sysname follows a linear control flow: preparation of MAS instances, a user-defined configuration specifies how a MAS is instantiated and executed, execution traces are collected during runtime, and post-hoc processing transforms these traces into interpretable metrics and summaries.
The workflow consists of five core components:

\para{MAS instances preparation.}
To use \sysname, users first need to prepare MAS instances to be evaluated. The details of the preparation process and the supported integration modes are described in \autoref{subsec: Integrating MAS Instances}.

\para{Configuration.}
Based on the prepared MAS instances, users specify the evaluation setup, including which input sources, the number and configuration of agent instances, and whether external tool access is enabled\footnote{Currently, \sysname only supports the adjustment of a few parameters, such as model choice and tool usage.}.
This configuration is passed to the \emph{Runtime} component for system instantiation and execution.

\para{Runtime.}
Based on the provided configuration, \emph{Runtime} component orchestrates the execution of the MAS instances.
Task inputs are continuously fed into the runtime, triggering agent interactions, tool invocations, and control-flow decisions as defined by the configuration.

\para{Observation.}
During execution, the \emph{Observation} component monitors system behavior through function-call hooks or sampling-based instrumentation.
Built on top of OpenTelemetry~\cite{opentelemetry} and psutil~\cite{psutil}, it records both default execution metrics (e.g., latency, token usage; see \autoref{subsec: otel template}) and additional signals specified in the configuration.
Collected traces are forwarded, either online or offline, to the \emph{Post-processing} component.

\para{Post-processing.}
The \emph{Post-processing} component aggregates and analyzes execution traces to make MAS behavior inspectable (e.g., CPU, call graph; see \autoref{subsec: Details of Post Processing Component}).
These summaries enable users to explore execution trajectories, compare configurations across runs, and identify performance bottlenecks and sources of instability.

\subsubsection{MAS instances preparation.}
\label{subsec: Integrating MAS Instances}

\begin{figure}[t]
\centering
\includegraphics[width=0.7\columnwidth]{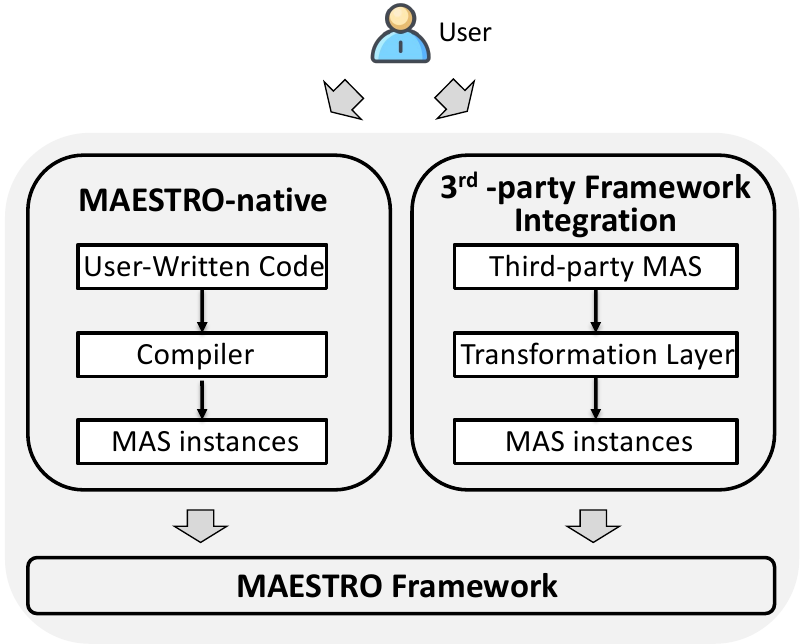}
\caption{Two ways to prepare MAS instances for \sysname, note that \sysname ships with a set of built-in MAS instances that can be used and compared directly.}
\label{fig: repo}
\end{figure}

Before performing any evaluation, MAS instances must be integrated into \sysname. As illustrated in \autoref{fig: repo}, users can prepare MAS instances in two ways:\footnote{At present, \sysname supports only \emph{pre-defined} MAS instances.}
\begin{itemize}[leftmargin=*]
  \item \textbf{\sysname-native}.
  Users can implement MAS instances directly using \sysname's native specification language and configuration interfaces. 
  This mode leverages compile-based techniques~\cite{cornacchia2025dmas} to automatically generate executable instances from high-level descriptions, ensuring optimal compatibility.
  This mode reduces manual coding effort by generating reusable scaffolding and integration code for common components.
\item \textbf{Third-party framework integration}.
  Through \sysname's transformation layer, users can build MAS instances in their preferred agent frameworks (e.g., ADK, LangGraph, AutoGen) or import existing open-source implementations, and connect them to \sysname for evaluation.
  The transformation layer provides a set of adapters that map framework-specific components onto \sysname's standard interfaces, exposing unified entry points for configuration, execution, and telemetry collection.
\end{itemize}
\sysname contributors can also use these two integration modes to add new built-in MAS instances to the framework. Currently, \sysname has 12 built-in MAS instances (described in detail in \autoref{subsec:example-studied}), allowing users to perform evaluations and comparisons directly without additional integration effort.

\subsection{MAS instances taxonomy}
\label{subsec:mas-taxonomy}

A well-designed benchmark should cover a broad (\textbf{O1}) and representative (\textbf{O2}) range of system configurations and use cases; otherwise, conclusions may overfit to a narrow slice of the MAS design space and fail to generalize. 
To enable systematic coverage and controlled comparisons, we characterize each MAS instance using a small set of well-defined dimensions. 
Specifically, we describe each instance along the following axes: application field, framework, interaction pattern, and data specification. 
These dimensions together capture the primary sources of variation in modern MAS deployments.

\para{Application field.} The high-level domain that the MAS instance targets, which may influence task complexity, required agent capabilities, and evaluation criteria. Common fields include question answering, creative generation, finance, and others.

\para{Framework.} The underlying multi-agent framework used to implement the MAS instance, which may affect agent orchestration, communication protocols, and tool integration. General frameworks include AutoGen~\cite{wu2024autogen}, ADK~\cite{adk2025}, LangGraph~\cite{langgraph}, and others.

\para{Interaction pattern.} The specific configuration of agents within the MAS instance, including the number of agents, the number of tools, and the cooperation type. Specifically, cooperation types include:
\begin{itemize}[leftmargin=*]
    \item \textit{Planning}. There is a dedicated planning agent that decomposes the task into subtasks and assigns them to other agents.
    \item \textit{Coordination}. Agents coordinate their actions through explicit communication.
    \item \textit{Debate}. Agents evaluate and compare candidate solutions to reach a consensus.
    \item \textit{Correction}. Agents collaboratively refine and improve a specific solution through iterative feedback.
\end{itemize}
These interaction patterns could affect the overall system dynamics and performance.

\para{Data specification.} The concrete input-output format and ground truth used to instantiate the MAS instance, which may influence task complexity, communication pattern, and evaluation criteria. 
The data specification could be divided into input and output.
\begin{itemize}
    \item \textit{Input.} A task is typically instantiated via a system prompt defining its core objectives and constraints.
    For tasks requiring open-ended exploration, the configuration phase may also incorporate external information retrieved through auxiliary tools, such as web search engines or private databases, as supplemental inputs. 
    Collectively, we define these structured inputs and retrieved data as \emph{artifacts}.
    \item \textit{Output.} According to whether the output has a determined ground truth, the output could be divided into \emph{Open-End} and \emph{Closed-Form}. 
\end{itemize}

\subsection{MAS example suites studied}
\label{subsec:example-studied}

\begin{table*}[t]
\centering
\caption{Selected MAS examples overview. The ``Suite'' column indicates membership, \textbf{F}: Full Suite; \textbf{A}: Architecture Suite.
}
\label{tab:example-overview}
\setlength{\tabcolsep}{6pt}
\renewcommand{\arraystretch}{1.15}

\begin{tabularx}{\textwidth}{
  >{\centering\arraybackslash}X
  >{\centering\arraybackslash}c
  >{\centering\arraybackslash}c
  >{\centering\arraybackslash}c
  >{\centering\arraybackslash}c
  >{\centering\arraybackslash}c
  >{\centering\arraybackslash}c
  >{\centering\arraybackslash}c
  >{\centering\arraybackslash}c
}
\toprule
\multirow{2}{*}{\textbf{Example}} &
\multirow{2}{*}{\textbf{App. Field}} &
\multirow{2}{*}{\textbf{Framework}} &
\multicolumn{3}{c}{\textbf{Interaction}} &
\multicolumn{2}{c}{\textbf{Data Spec.}} &
\multirow{2}{*}{\textbf{Suite}} \\
\cmidrule(lr){4-6}\cmidrule(lr){7-8}
& & &
\textbf{Type} & \textbf{\#Agt} & \textbf{\#Tool} &
\textbf{In} & \textbf{Out} &
\\
\midrule
Fin. Analyzer~\cite{lastmile2024mcpfinancial} & Finance & MCP-Agent & Correct & 6 & 1 & Artifacts & Opn-End & F \\
Img. Scr.~\cite{adksamples} & Creativity & ADK & Debate & 4 & 2 & Artifacts & Cls-Form & F \\
Marketing~\cite{adksamples} & Marketing & ADK & Coord. & 4 & 1 & Artifacts & Opn-End & F \\
Brand SEO~\cite{adksamples} & Marketing & ADK & Coord. & 4 & 10 & Artifacts & Opn-End & F \\
Content Creat.~\cite{a2asamples} & Creativity & ADK & Plan. & 4 & 1 & Artifacts & Opn-End & F \\
Mag.-One~\cite{fourney2024magentic-one} & Cross-domain & Autogen & Plan & 4 & 0 & Artifacts & Opn-End & F \\
Stock Res.~\cite{autogen} & Finance & Autogen & Coord. & 4 & 2 & Artifacts & Opn-End & F \\
Travel Plan.~\cite{autogensamples} & Travel & Autogen & Coord. & 4 & 0 & Artifacts & Opn-End & F \\
ToT~\cite{yao2023tree} & Cross-domain & LangGraph & Debate & 3 & 0 & Artifacts & Cls-Form & F \\
CRAG~\cite{yan2024corrective} & Cross-domain & LangGraph & Coord. & 5 & 2 & Datasets & Opn-End & F,A \\
Plan\&Exec.~\cite{wang2023plan} & Cross-domain & LangGraph & Plan & 3 & 1 & Datasets & Opn-End & F,A \\
LATS~\cite{zhou2023language} & Cross-domain & LangGraph & Plan & 3 & 1 & Datasets & Opn-End & F,A \\
\bottomrule
\end{tabularx}
\end{table*}

We carefully select 12 representative MAS instances to serve as the pre-defined evaluation set in \sysname.
These instances are designed to provide sufficient coverage of common MAS configurations and use cases (\textbf{O1, O2}), and to act as a baseline for subsequent studies.
As summarized in \autoref{tab:example-overview}, the selected instances are chosen according to the following criteria:

\begin{itemize}[leftmargin=*]
    \item \textbf{Framework diversity (O2, O3)}: We include examples implemented using different popular MAS frameworks, such as \emph{MCP-Agent}, \emph{LangGraph}, \emph{ADK}, and \emph{Autogen}, to capture a wide range of design patterns and interaction paradigms.
    \item \textbf{Official sources (O2)}: We collect examples that are provided in the official example repositories or tutorials of these frameworks, ensuring that they reflect best practices and standard usage patterns.
    \item \textbf{Domain variety (O2)}: We select examples that cover diverse application domains, including question answering, planning, creative writing, marketing strategy, and so on, to evaluate MAS performance across different application scenarios.
    \item \textbf{Interaction diversity (O1)}: We prioritize examples that exhibit varied interaction patterns among agents, such as cooperative problem solving, debate-style discussions, and role-based collaborations, to assess how different interaction styles affect MAS behavior.
\end{itemize}

As a prerequisite for meaningful analytical post-processing, MAS instances must be grouped into coherent categories that share relevant characteristics. 
Such grouping enables comparative analysis across multiple configurations, ensuring that observed behaviors reflect systematic trends rather than ad hoc artifacts of individual runs.
To demonstrate the analytical capabilities of \sysname, we derive two evaluation suites, each designed to surface distinct system-level insights.

\begin{itemize}[leftmargin=*]
    \item \textbf{Full-suite (F)}: This suite includes all selected MAS examples. We treat this suite as a representative subset of real-world MAS deployments, and use it to study the overall performance and behavior of LLM-based MAS in realistic settings.
    \item \textbf{Architecture-focused suite (A)}: This suite includes three representative MAS examples that implement different representative multi-agent architectures but solve the same set of tasks. 
    This suite is used to study the impact of agent architectures on MAS behavior.
\end{itemize}

\begin{figure}[t]
\centering
\includegraphics[width=0.95\columnwidth]{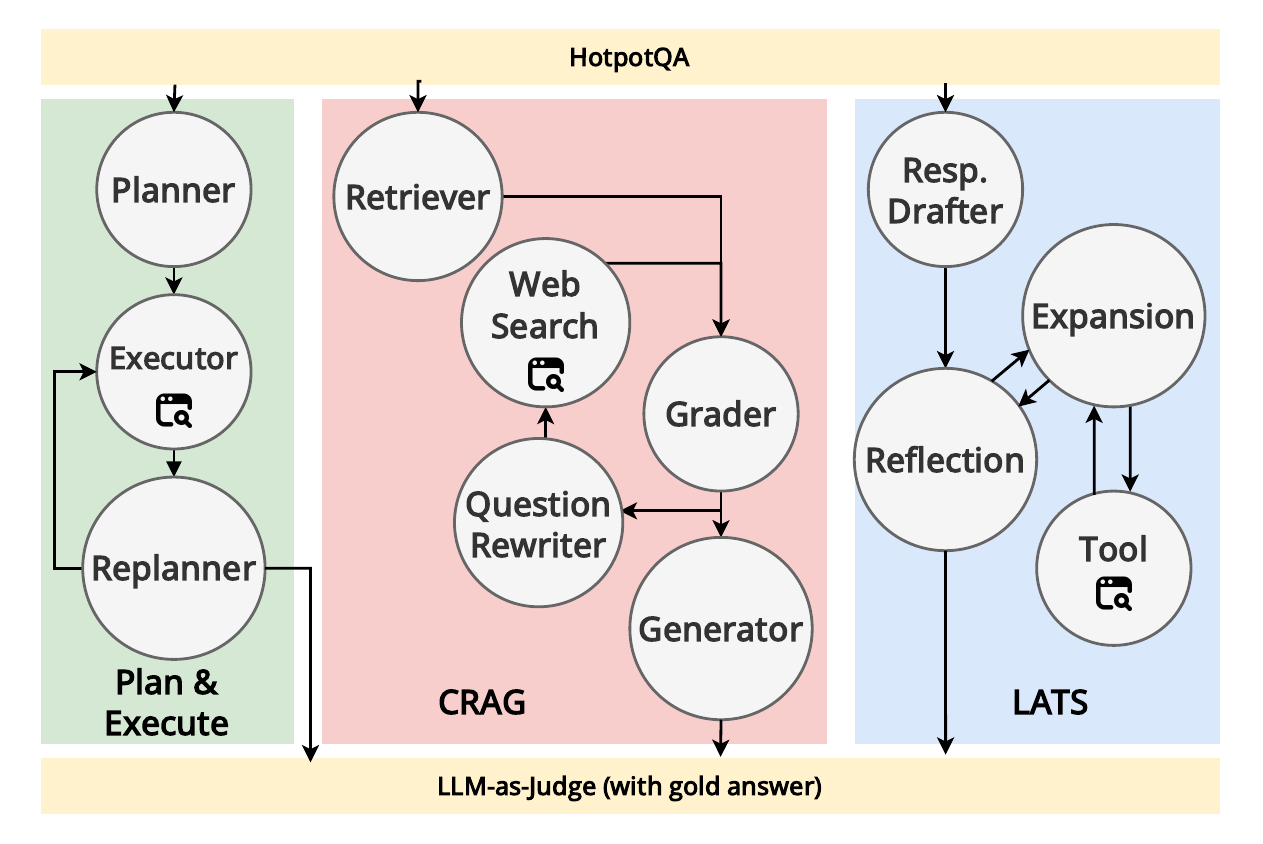}
\caption{Solving the same given tasks with 3 different MAS architectures.}
\label{fig:arch-comps}
\end{figure}

In the architecture-focused suite, we select three representative MAS architectures: CRAG (Corrective RAG)~\cite{yan2024corrective}, Plan\&Execute~\cite{wang2023plan}, and LATS (Language Agent Tree Search)~\cite{zhou2023language}. 
They are designed as general-purpose agent architectures that can operate across tasks without being tightly coupled to specific applications. 
However, their design goals differ. 
As shown in ~\autoref{fig:arch-comps}, CRAG is optimized for retrieval-centric workloads, whereas LATS and Plan-and-Execute target more general problem-solving settings, employing tree-search–based divide-and-conquer and greedy iterative refinement strategies, respectively.
Such variety in the task-solver architectures enables the following comparative studies.

In the following section, we present case studies and analyses using \sysname, organized around these two MAS example suites.\section{Case studies}
\label{sec: Case studies}

To demonstrate how researchers can benefit from \sysname, we conduct a series of case studies that illustrate the types of insights enabled by its fine-grained telemetry.
We organize our evaluation into two complementary sets of case studies:
\begin{itemize}[leftmargin=*]
    \item \textbf{General system-level analysis.}
    In \autoref{sec:mas-overall-res-consumption} and \autoref{sec:mas-graph-stability}, we examine system-level metrics that are not specific to MAS, but instead provide a familiar baseline for reasoning about performance, resource consumption, and reliability, analogous to evaluations in traditional systems.
    \item \textbf{Application- and semantics-aware analysis.}
    Using Evaluation Suite~2, we investigate how different MAS solver architectures affect cost, latency, and accuracy (\autoref{sec:mas-arch-impact}, \autoref{sec:mas-model-impact}, \autoref{sec:mas-failure-attribution}, and \autoref{sec:mas-tool-impact}).
    This analysis explores whether architectural choices and structural optimizations lead to consistent performance trade-offs, in a manner analogous to ablation studies.
\end{itemize}

\subsection{Methodology}
\label{sec:methodology}
\para{Inputs.}
For each MAS instance, we generate evaluation inputs using one of the following three approaches:
\noindent\begin{itemize}[leftmargin=*]
    \item \textbf{Naive artifacts.} Direct reuse of input prompts provided in the README files of official example repositories.
    \item \textbf{Public datasets.} Inputs drawn from publicly available benchmark datasets aligned with the task domain of the MAS instance (e.g., question-answering datasets for QA-oriented agents~\cite{yang2018hotpotqa}).
    \item \textbf{Synthetic inputs.} LLM-generated prompts that enable controlled variation and increased input diversity.
\end{itemize}

\para{Setup.} For each MAS instance, we conduct at least 20 independent runs to characterize execution behavior. 
Each run consists of submitting a single user-level task input to the MAS (e.g., a single ``write a blog post'' prompt for \emph{Content Creation}).
For MAS instances that require human-in-the-loop interaction, user responses are simulated using an LLM-as-user approach, where a designated LLM (\texttt{gemini-2.5-flash} in our current setup) generates replies conditioned on the MAS outputs.
To prevent non-terminating execution, each run is capped at 10 minutes.
For the architecture-focused suite, LLM responses are additionally limited to a maximum of 8{,}192 tokens.
As for the external tool usage mentioned in \autoref{tab:example-overview}, we use Tavily or Google Search~\cite{googleSearchTool} for web search, and use Google \texttt{imagen-3.0-generate-002}~\cite{googleImageGenTool} model for image generation.
When correctness evaluation is required, we employ \texttt{gpt-4o-mini} as an LLM-as-judge.
To evaluate the impact of different backbone models, we vary the underlying LLM across several configurations: Gemini-2.0-Flash-Lite (Ge20FL), Gemini-2.5-Flash-Lite (Ge25FL), Gemini-2.5-Flash (Ge25F), GPT-4o-mini (G4oM), GPT-5-mini (G5M), and GPT-5-nano (G5N).

\subsection{What are the systems usage patterns and implications?}
\label{sec:mas-overall-res-consumption}
Resource consumption such as CPU, memory, and network usage are important factors to consider when deploying MAS in real-world systems. Understanding the resource usage patterns of MAS can help optimize their performance and scalability. In this subsection, we analyze the resource consumption of MAS and investigate the factors that influence their usage patterns.

\begin{figure}[t]
\centering
\includegraphics[width=1\columnwidth]{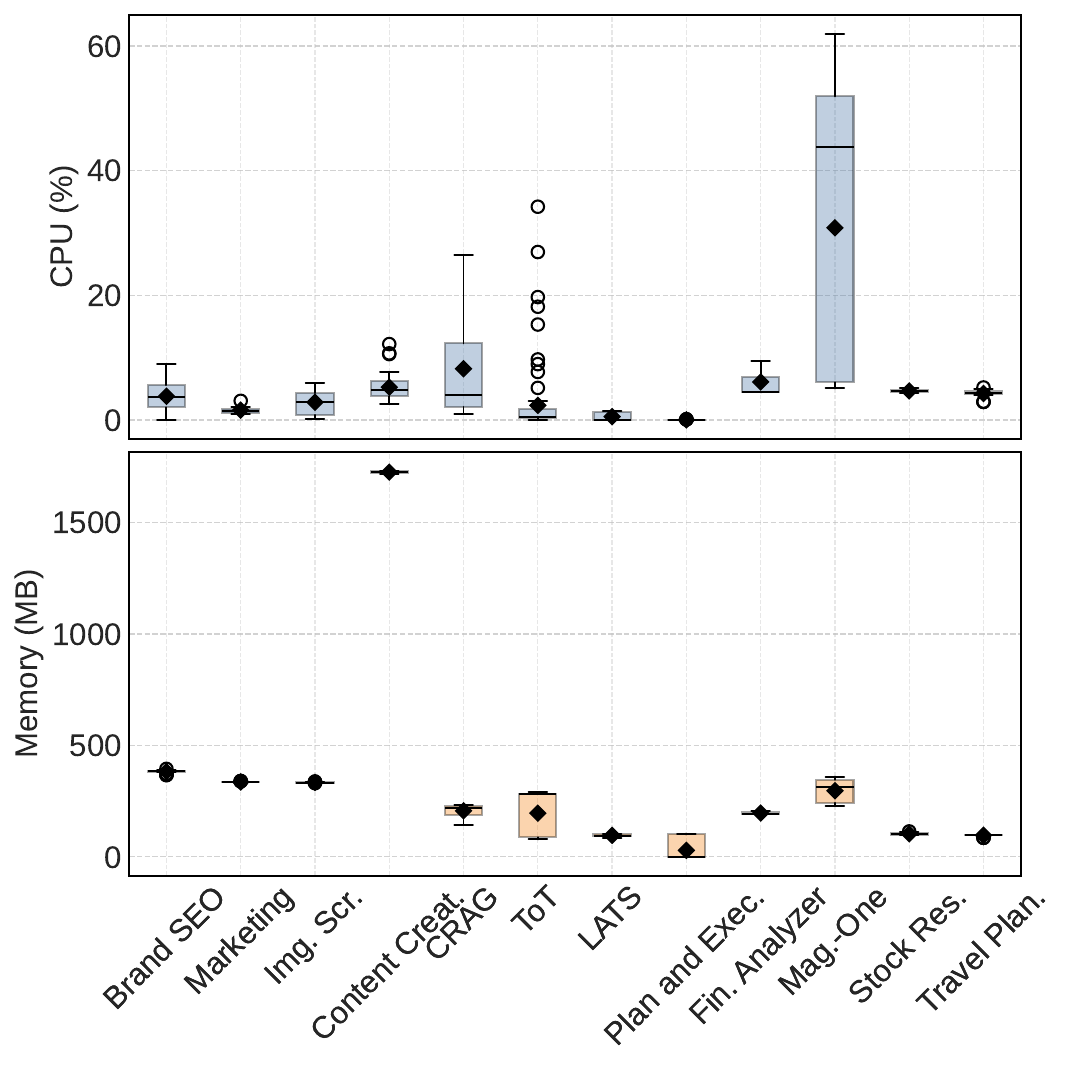}
\caption{CPU and memory usage across different examples.}
\label{fig: cpu_mem_by_examples}
\end{figure}

\para{Per-task CPU and memory footprints are modest and bounded in our setup.}
\autoref{fig: cpu_mem_by_examples} reports per-run CPU and memory usage across the 12 MAS examples under our runtime configuration.
For CPU, the maximum observed utilization reaches 61.9\%, while the boxplot distributions for most examples remain substantially below this peak, suggesting that these workloads typically do not require sustained heavy local compute.
For memory, excluding the Content Creation example which peaks at 1726.8\,MB, the average memory usage across examples is 200.2\,MB, placing most single-task executions in a sub-GB regime in our measurements.
The Content Creation example’s higher memory footprint stems from its distributed design, where each agent runs in a separate process, increasing the aggregate resident memory. 
We leave a broader study of distributed MAS deployments and their resource trade-offs to future work.
We also observe framework-dependent memory patterns; for example, examples implemented with \textit{ADK} tend to exhibit higher memory footprints in our setup.

\begin{figure*}[t]
\centering
\includegraphics[width=1.98\columnwidth]{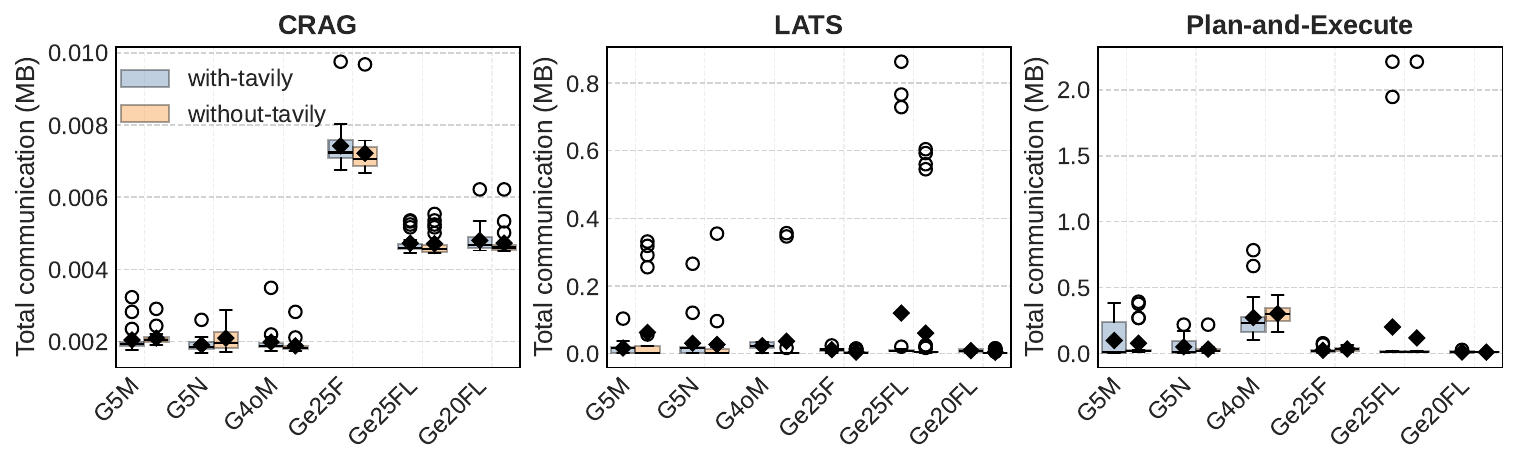}
\caption{Communication usage across different architectures.}
\label{fig: comm_with_without}
\end{figure*}

\para{Per-task communication volume is in the MB scale and varies by architecture and model.}
We further measure total communication volume across architectures with and without tools (\autoref{fig: comm_with_without}).
Across all tested configurations, the observed communication volumes stay within the MB scale (with most cases within a few MB), indicating that, per task, network payload is typically small compared to CPU and memory footprints.
It can be observed that tool usage has a minor impact on communication volume. 
Model choice also interacts with architecture; for example, in CRAG, Gemini-family models show larger communication volumes than GPT-family models in our measurements.

\rqbox{In our setup, most single-task executions stay within a sub-GB memory regime and bounded CPU utilization, while per-request communication remains at the MB scale.}

\begin{figure*}[t]
\centering
\includegraphics[width=2\columnwidth]{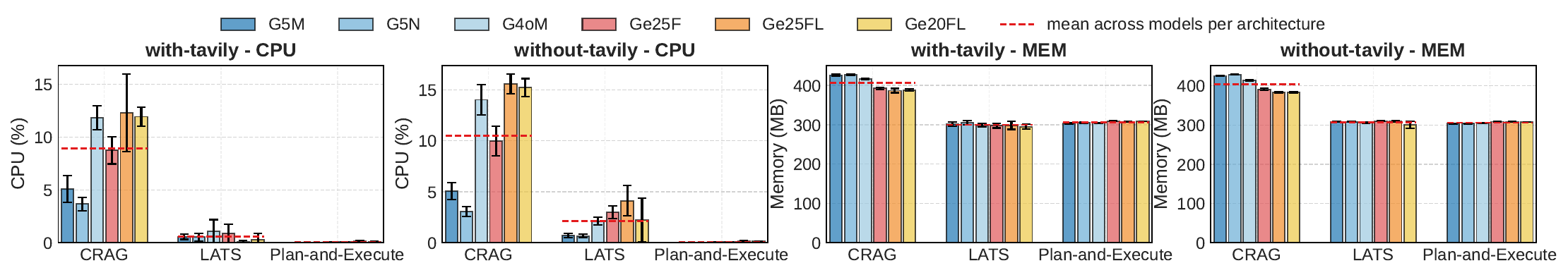}
\caption{CPU and memory usage across different architectures.}
\label{fig: cpu_mem_by_example_grouped}
\end{figure*}

\para{CPU and memory usage patterns are architecture-dependent.}
\autoref{fig: cpu_mem_by_example_grouped} illustrates the CPU and memory usage grouped by architecture. Consistent with the communication patterns observed in \autoref{fig: comm_with_without}, resource consumption is highly architecture-dependent. CRAG exhibits the highest resource footprint, with an average CPU usage of 9.7\% and memory usage of 405.3MB (averaged over all model and tool configurations). This is followed by Language Agent Tree Search (1.36\% CPU) and Plan-and-Execute (0.07\% CPU). We further observe that while model choice influences CPU load, it has a negligible impact on memory. Surprisingly, enabling tools reduces global average CPU usage by 3.1\% and memory by 4.8MB. This phenomenon could be explained by the fact that tool usage could help reduce the number of LLM calls, which are CPU and memory-intensive.

\rqbox{Architecture dominates resource patterns, while model choice introduces smaller shifts.}

\subsection{How stable are MAS call graphs, and what factors influence their variability?}
\label{sec:mas-graph-stability}

A distinguishing characteristic of LLM-based MAS, in contrast to traditional systems like microservices, is the inherent stochasticity of their execution behavior. In microservice architectures, call graph variability is widely used as an indicator of anomalous executions and edge cases. By extending this concept to MAS, one can leverage variability to sample a diverse and representative set of execution traces. However, a foundational stability analysis is a prerequisite for such methodologies. From a reproducibility perspective, higher call graph similarity across repeated runs implies stronger run-to-run consistency in agent interactions, and thus more reproducible MAS executions. Therefore, quantifying the stability of agent interactions across different runs and identifying the factors influencing their variability are crucial for designing robust and reproducible MAS.

We use two metrics to measure the call graph similarity: Jaccard similarity and Largest Common Sequence (LCS) similarity. These metrics capture different aspects of the call graph structure and provide insights into the agent interactions.

\begin{itemize}
    \item \textbf{Jaccard similarity (edge-set overlap)}:
    For each run $i$, we construct a directed call graph $G_i$ and denote by $E_i$ its (unweighted) edge set.
    For runs $i,j$, we compute
    \[
        J(E_i,E_j)=\frac{|E_i\cap E_j|}{|E_i\cup E_j|},
    \]
    with the convention $J(E_i,E_j)=0$ when $E_i\cup E_j=\emptyset$.
    This captures whether the same interaction edges appear at least once, regardless of frequency.

    \item \textbf{LCS (order consistency)}:
    For each run $i$, we linearize the calls into an ordered edge sequence $S_i$.
    Let $\mathrm{LCSlen}(S_i,S_j)$ denote the length of the longest common subsequence between $S_i$ and $S_j$.
    We define the normalized LCS similarity as
    \[
        \mathrm{LCS}(S_i,S_j)=\frac{\mathrm{LCSlen}(S_i,S_j)}{\max(|S_i|,|S_j|)},
    \]
    with the convention that two empty sequences yield $1$ and one empty sequence yields $0$.
    This measures the consistency of interaction order.

\end{itemize}

To summarize similarity at different granularities (e.g., per example, per model, or per experimental condition such as tool-on vs.\ tool-off), we first partition runs into groups according to the dimension of interest. For a group with $n$ runs, we compute similarity values for all unordered run pairs $(i,j)$ with $1\le i<j\le n$. For each pair, we compute $J(E_i,E_j)$ and $\mathrm{LCS}(S_i,S_j)$. We define the \emph{pairwise average similarity} for a group as the mean of these pairwise values over all unordered run pairs in that same group. If $n<2$, we set the pairwise average similarity to $0$ (no pairwise comparisons are available).

\begin{figure}[t]
\centering
\includegraphics[width=1\columnwidth]{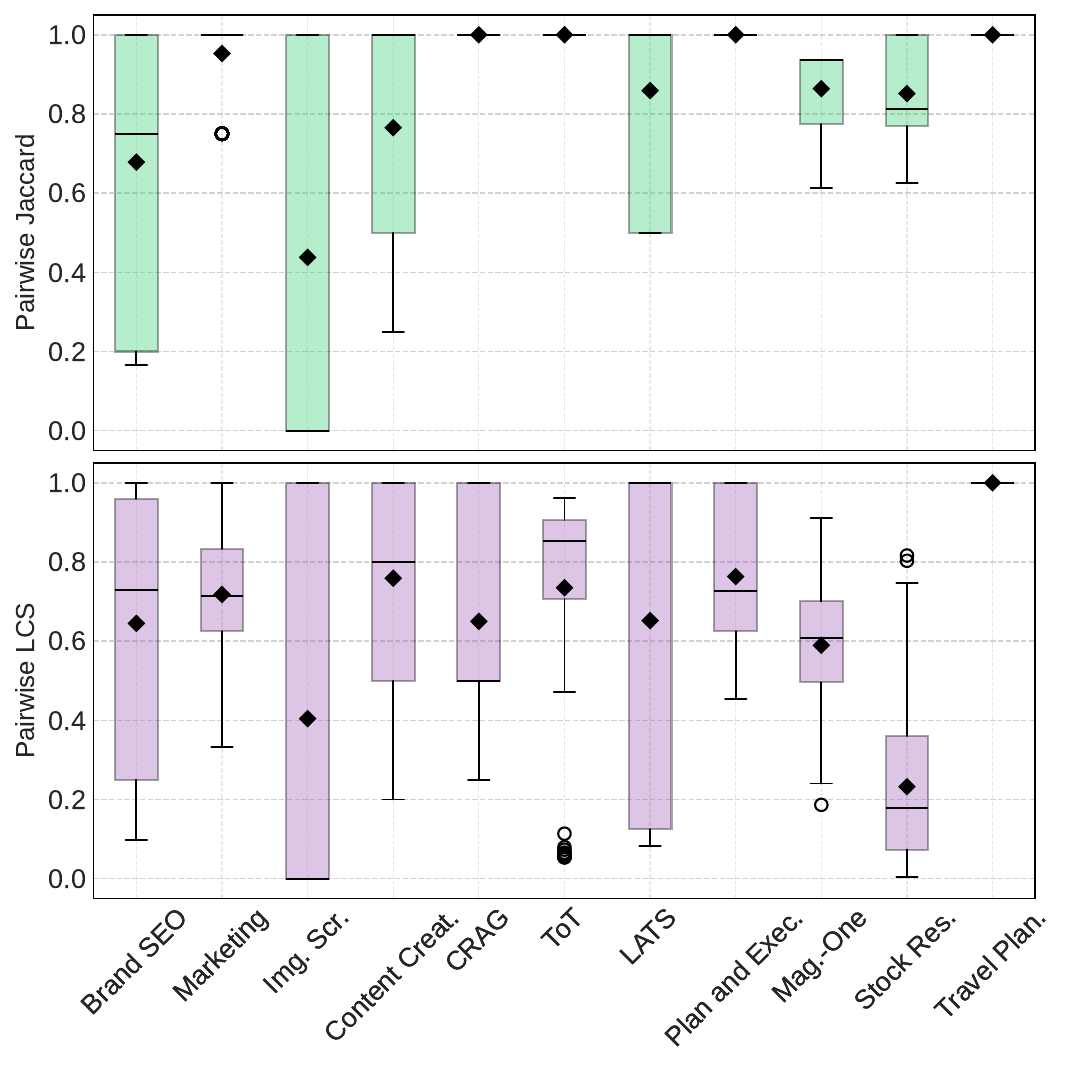}
\caption{Cross-model call graph similarity (Jaccard Similarity: Order-agnostic overlap of agent interactions, LCS Similarity: Order-aware similarity of execution traces).
}
\label{fig:model-sim-suite1}
\end{figure}

\para{MAS execution exhibits structural stability but sequential variance.} 
We first examine the stability of call graphs across repeated runs of the same example. 
\autoref{fig:model-sim-suite1} presents the intra-example average pairwise similarity for the Full Suite. 
We observe that across all cases there exists high Jaccard similarities (average 0.86 across all examples), indicating that the \emph{set of agent-to-agent interactions} remains robust against execution variance. 
In contrast, LCS similarity is moderate (average 0.65), suggesting that the \emph{sequence of agent calls} fluctuates significantly across runs. 
Notably, examples like CRAG and Tree-of-Thoughts demonstrate high Jaccard but low LCS scores, confirming that while the participating agents and their connections remain consistent, the temporal order of their interactions is highly dynamic. 
A distinct exception is the travel-planning example, which employs the \textit{RoundRobinGroupChat} mechanism from \emph{Autogen}; this enforces a deterministic execution order, resulting in perfect stability (1.0) for both metrics.

\rqbox{Across runs, MAS call graphs are largely stable in which agent-to-agent interactions occur, but often unstable in the order those interactions unfold; consequently, reproducibility is stronger at the interaction-structure level than at the execution-order level.}

\begin{figure}[t]
\centering
\includegraphics[width=1\columnwidth]{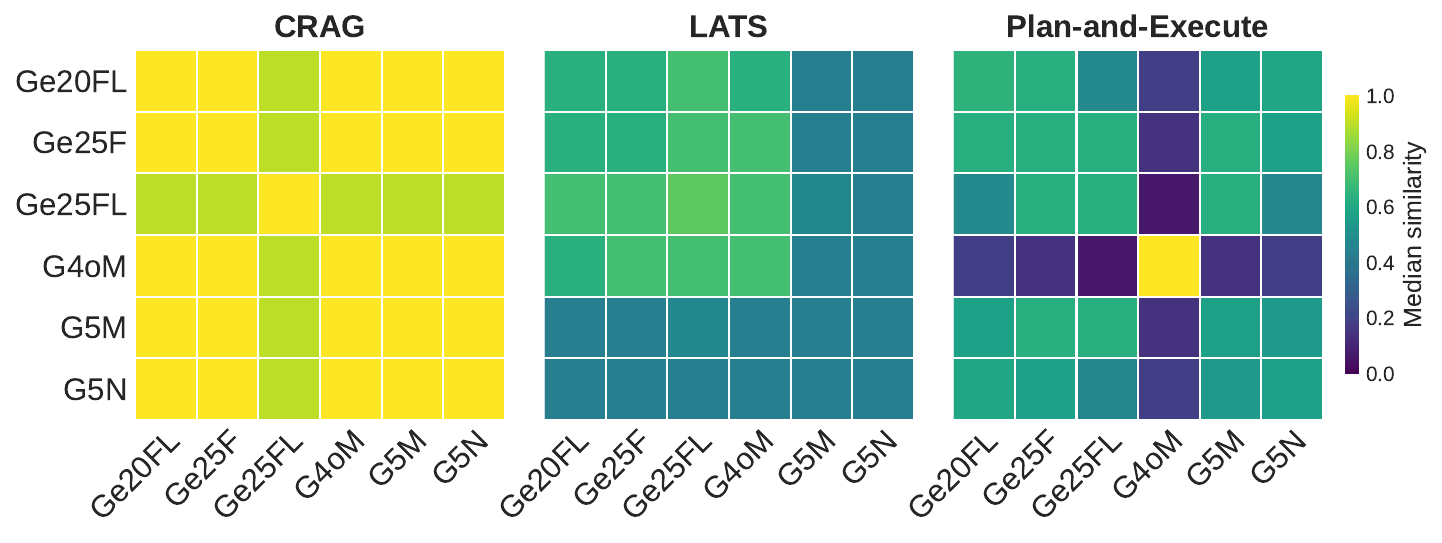}
\caption{Cross-model call graph LCS similarity.}
\label{fig:model-sim-suite2}
\end{figure}

\para{Architecture determines stability, while model impact is architecture-specific.}
We further investigate the factors influencing call graph similarity by comparing execution patterns across different models and architectures. \autoref{fig:model-sim-suite2} presents the cross-model LCS call graph similarity heatmap, where each sub-figure corresponds to an architecture and each cell represents the median pairwise similarity between two models. We observe distinct stability profiles across architectures: CRAG exhibits extremely high consistency (average similarity 0.97 across all model pairs), whereas Language Agent Tree Search (0.54) and Plan-and-Execute (0.47) show significantly more variation. Furthermore, the impact of model choice is architecture-dependent. In CRAG, most models share identical call graphs, with \texttt{gemini-2.5-flash-lite} being the sole outlier. In Language Agent Tree Search, all models produce similar call graphs. Conversely, for Plan-and-Execute, \texttt{gpt-4o-mini} diverges significantly from other models by showing low similarity to them, yet it remains highly self-consistent across its own runs, while the remaining models tend to resemble one another.

\rqbox{MAS architecture dominates the call graph similarity, with model choice having different effects depending on the architecture.}

We then move to application-level metrics, including cost, task duration, and accuracy. 
Due to the inherent non-determinism of LLMs, accuracy can vary across runs; ensuring result quality therefore often correlates with increased cost and longer execution time. 
\sysname enables systematic analysis of such behavior despite the chaotic nature of LLM-driven execution. 
To ensure fair comparison across configurations, the following evaluation focuses exclusively on the Architecture Suite, which supports finer-grained analysis.

\subsection{How do different agent architectures affect task performance and stability?}
\label{sec:mas-arch-impact}

\begin{figure*}
\centering
    \begin{subfigure}[t]{0.71\columnwidth}
    \includegraphics[width=\columnwidth]{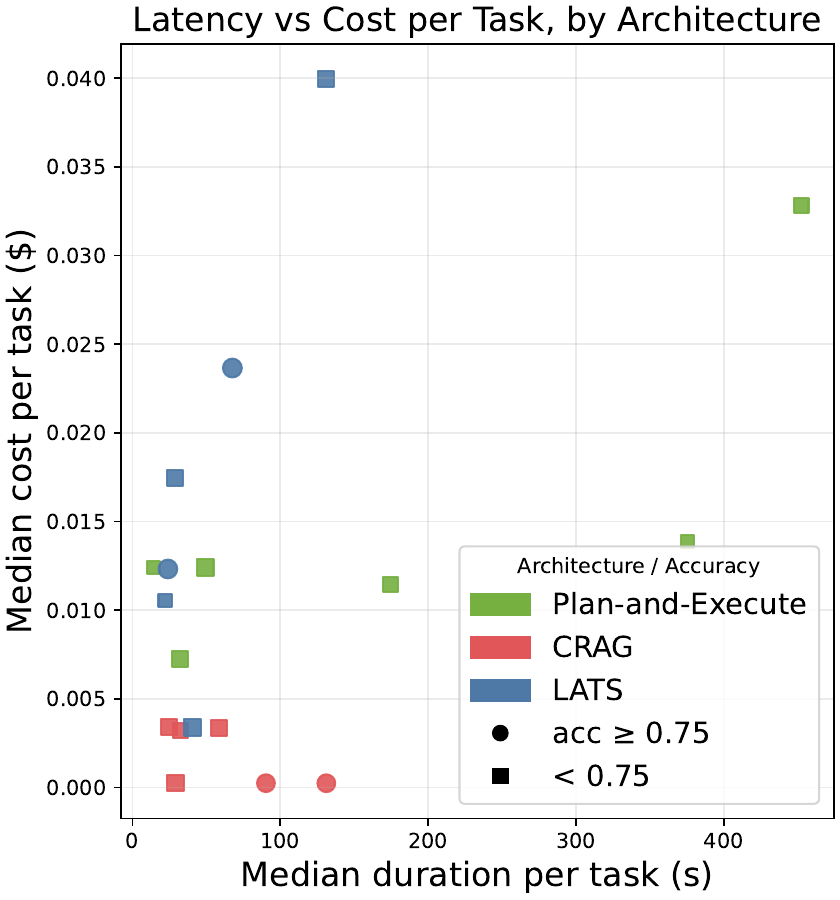}
        \caption{Latency versus cost per task, grouped by agent architecture.
        Lower values on both axes indicate better performance.}
        \label{subfig:lat_cost_per_arch}
    \end{subfigure}
    \hfill
    \begin{subfigure}[t]{1.25\columnwidth}
        \includegraphics[width=\columnwidth]{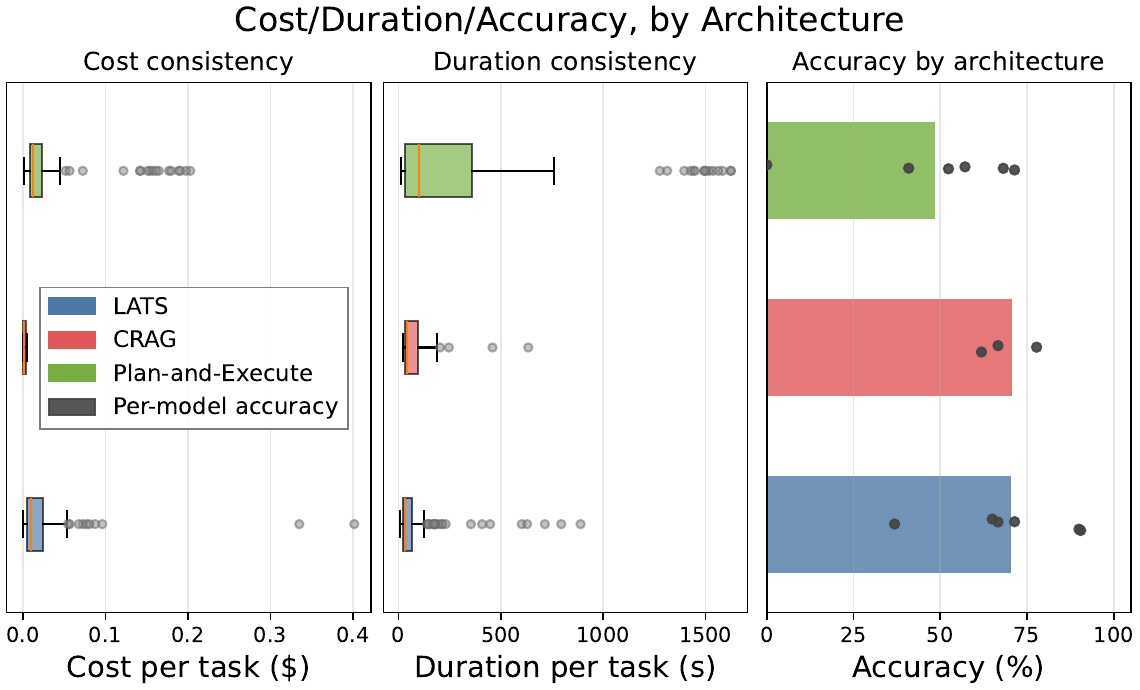}
        \caption{Distribution of cost, latency, and accuracy across agent architectures; specialized designs (e.g., CRAG) achieve stable accuracy at lower cost.}
        \label{subfig:const_per_arch}
    \end{subfigure}
    \caption{
    Resource cost versus accuracy across agent architectures.
    Task-specific designs such as CRAG achieve comparable accuracy with lower resource cost than more general architectures like LATS.
    }
    \label{fig:rq1}
\end{figure*}

More general-purpose solver architectures, designed to handle a wide range of complex tasks, tend to progress more cautiously. 
To ensure robustness, they often pause at each iteration to reflect on intermediate states. 
For example, Plan-and-Execute first decomposes the overall goal into a sequence of milestones and then solves each subtask incrementally. 
This approach helps the model maintain a comprehensive understanding of task context and often yields more reliable outcomes, but at the cost of increased execution time and resource consumption.

In contrast, when the task type is known in advance, a more specialized architecture can be employed. 
CRAG, for instance, is explicitly designed for retrieval-based workloads. Rather than exploring alternative reasoning paths, it prioritizes directly answering the query with minimal detours. 
This objective-driven design attempts to solve the task as early as possible, even with incomplete background information, trading exploration for efficiency. 
Such differences in design philosophy lead to substantial divergence in execution behavior across architectures.

\para{Specialized solver minimizes resource consumption.}
As shown in \autoref{subfig:lat_cost_per_arch}, CRAG consistently occupies the lower-cost and lower-latency region across different model choices. 
In particular, CRAG achieves a median cost of \$0.0010 per task, which is more than an order of magnitude lower than both Plan-and-Execute (median \$0.0126) and LATS (median \$0.0101). 
CRAG also executes faster, with a median task duration of 42.8\,s, compared to 101.5\,s for Plan-and-Execute.

In contrast, Plan-and-Execute exhibits substantially higher variance in task duration (interquartile range 30.6--356.6\,s), reflecting the overhead introduced by iterative planning and execution. 
LATS achieves relatively low median latency (32.3\,s), but incurs higher resource cost overall.

\para{Accuracy degrades with increasing architectural complexity.}
Furthermore, \autoref{subfig:const_per_arch} shows that CRAG attains accuracy comparable to, and in some cases exceeding, more general architectures. 
CRAG achieves an average accuracy of 70.6\%, compared to 48.3\% for Plan-and-Execute, while also exhibiting lower variability across runs. 
These results indicate that task-specialized agent architectures can simultaneously reduce resource consumption and maintain strong task performance.

Notably, increased architectural complexity does not necessarily translate into higher accuracy and may even be detrimental. 
While it is tempting to introduce additional agents -- such as fact-checkers or verification stages -- to enforce desired behavior, such designs inevitably increase execution cost and prolong interaction histories. 
In our evaluation, Plan-and-Execute spends substantially more time reasoning over tasks yet achieves lower accuracy (average 48.3\% vs.\ 70.6\% for CRAG), despite incurring significantly higher execution cost. 
This behavior aligns with prior findings that model performance degrades as interaction histories grow longer, due to diminishing attention to earlier context and error accumulation in extended reasoning chains~\cite{liu2024lost}.

\rqbox{More general agent architectures consume more resources and do not consistently improve accuracy.}

\subsection{How does model choice affect MAS behavior?}
\label{sec:mas-model-impact}

A natural assumption in LLM-based MAS design is that upgrading the underlying model should improve system performance. 
Intuitively, scaling to more capable models is expected to increase cost while yielding higher accuracy. 
However, our experimental results challenge this assumption. 
We find that stronger models do not necessarily incur substantially higher costs in practice, nor do they consistently lead to improved correctness. 
Instead, model choice affects MAS behavior in more nuanced and sometimes counterintuitive ways.

\para{Stronger models reduce iteration overhead rather than total cost.}
More capable models often complete subtasks with fewer iterations, reducing pathological behaviors such as repeated retries or prolonged refinement loops. 
However, these efficiency gains primarily offset higher per-token pricing rather than translating into lower overall cost. 
For example, \texttt{gpt-5-mini} and \texttt{gpt-5-nano} exhibit comparable mean cost per task (0.033 vs.\ 0.043), despite differences in model size, while \texttt{gpt-4o-mini} achieves substantially lower median cost (0.0034) than both. 
Similarly, execution latency is non-monotonic: \texttt{gpt-4o-mini} completes tasks faster (median 45.3\,s) than the larger 5-series models, whereas \texttt{gpt-5-nano} is slower than \texttt{gpt-5-mini} despite being nominally smaller. 
These results indicate that model choice influences iteration efficiency and tail behavior, but does not induce a clear cost hierarchy.

\para{Accuracy exhibits non-monotonic and unstable trends across models.}
We further observe no consistent relationship between model strength and task accuracy. 
While \texttt{gpt-5-mini} achieves the highest accuracy (median 81\%), weaker or similarly priced models do not follow a predictable trend: \texttt{gpt-5-nano} trails at 65\%, and \texttt{gpt-4o-mini} exhibits high median accuracy (71\%) but a substantially lower mean (48\%), indicating unstable behavior with heavy failure cases. 
Gemini models cluster around similar accuracy levels (approximately 66\%), with the 2.0-lite variant performing worse overall. 
These results suggest that MAS accuracy is highly sensitive to execution dynamics and variance amplification, rather than model capacity alone, and that upgrading the base model is insufficient to guarantee improved correctness.

\rqbox{Upgrading the base LLM does not reliably reduce cost or improve accuracy in MAS, as execution dynamics dominate model-level gains.}

\begin{figure}[t]
\centering
\includegraphics[width=0.95\columnwidth]{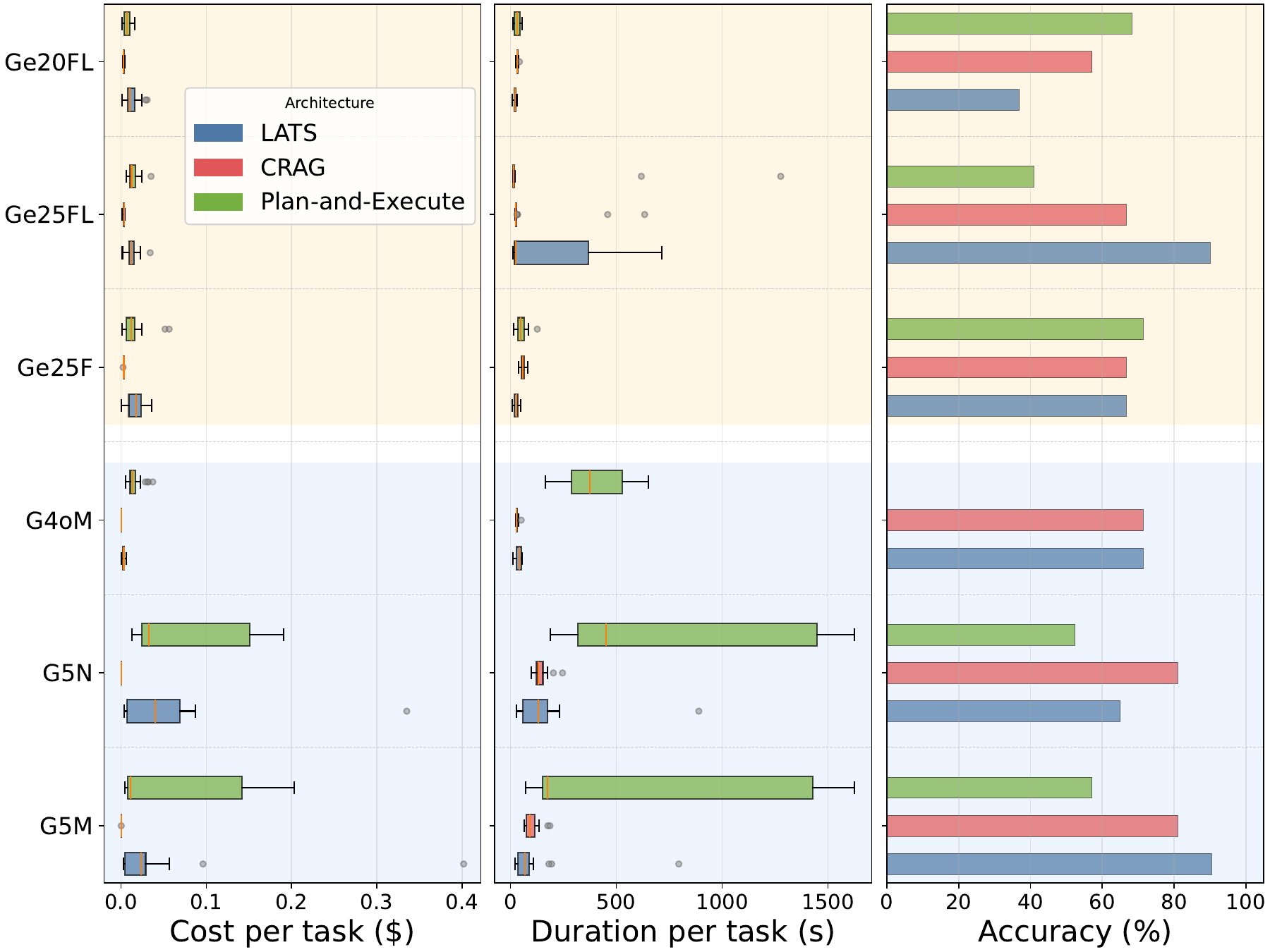}
\caption{Cost–duration–accuracy trade-offs across LLMs; efficiency improves for Gemini-family models, while accuracy shows no clear scaling trend.}
\label{fig:model-tradeoff}
\end{figure}

\subsection{What are the dominant failure modes in LLM-based multi-agent systems?}
\label{sec:mas-failure-attribution}

We find that most failures manifest as silent gray errors (75.17\% in \autoref{tab:failure_composition}), which do not trigger explicit system failures and are therefore not immediately visible to users. 
These errors only become apparent upon manual inspection of the output. 
Importantly, such failures are not system-level exceptions, but rather plausible-looking yet unusable responses. 
As a result, failure attribution in LLM-based MAS is particularly challenging, since erroneous executions often complete without emitting any hard error signals.

\begin{table}[t]
\centering
\caption{Global failure composition across all experiments.}
\label{tab:failure_composition}
\small
\begin{tabular}{l r}
\toprule
\textbf{Failure category} & \textbf{Percentage (\%)} \\
\midrule
Missing / underspecified output & 47.61 \\
Wrong fact / entity             & 27.66 \\
Empty prediction                & 15.96 \\
Exception                       & 6.38  \\
Timeout                         & 1.86  \\
Other                           & 0.54 \\
\midrule
\textbf{Silent semantic failures} & \textbf{75.17} \\
Explicit failures & 24.84 \\
\bottomrule
\end{tabular}
\end{table}

\begin{figure*}[t]
\centering
    \begin{subfigure}[t]{0.70\columnwidth}
        \includegraphics[width=\columnwidth]{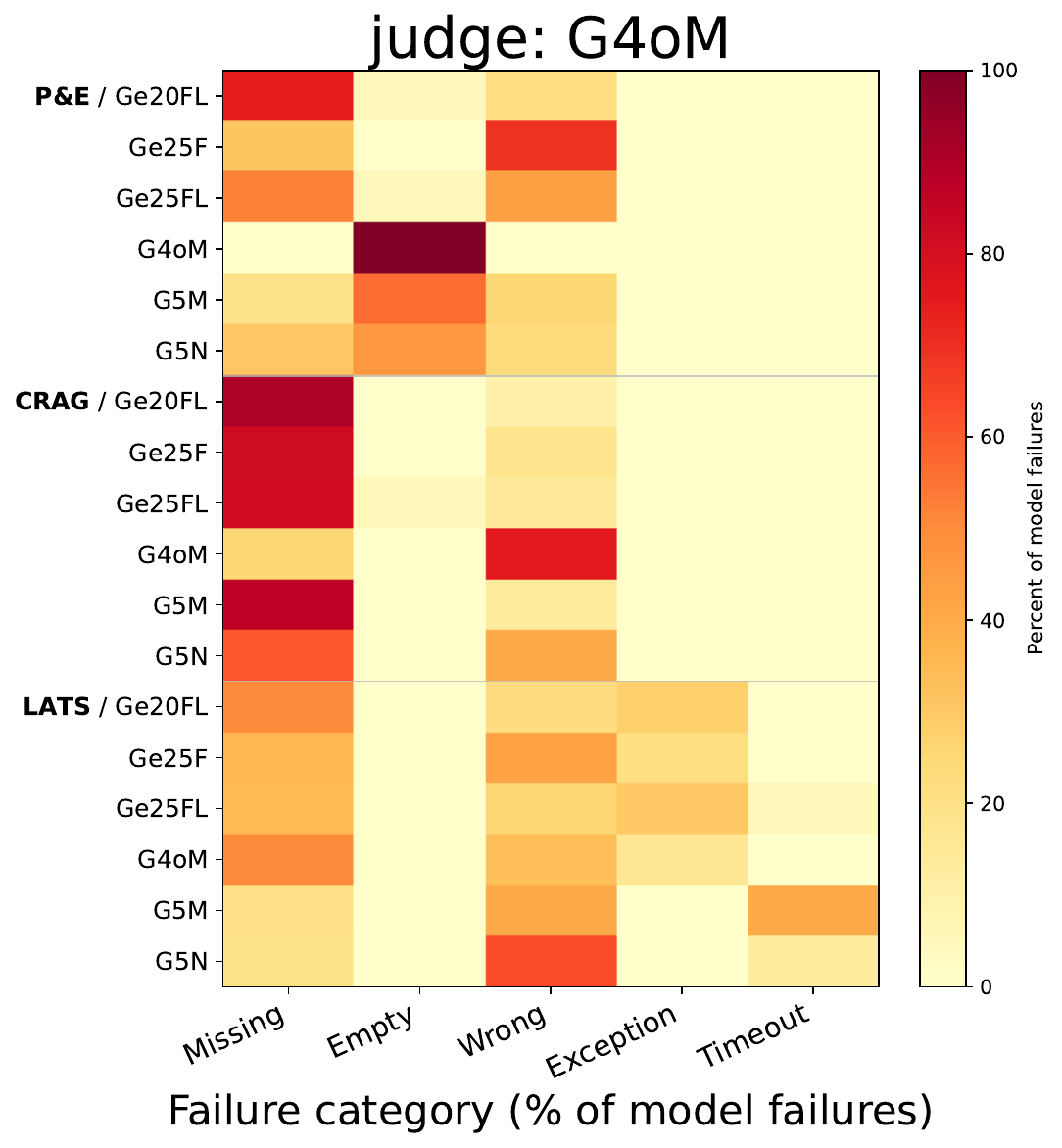}
        \caption{Online failure attribution via LLM-as-judge (GPT-4o-mini).}
        \label{subfig:failure_gpt4o}
    \end{subfigure}
    \hfill
    \begin{subfigure}[t]{1.26\columnwidth}
        \includegraphics[width=\columnwidth]{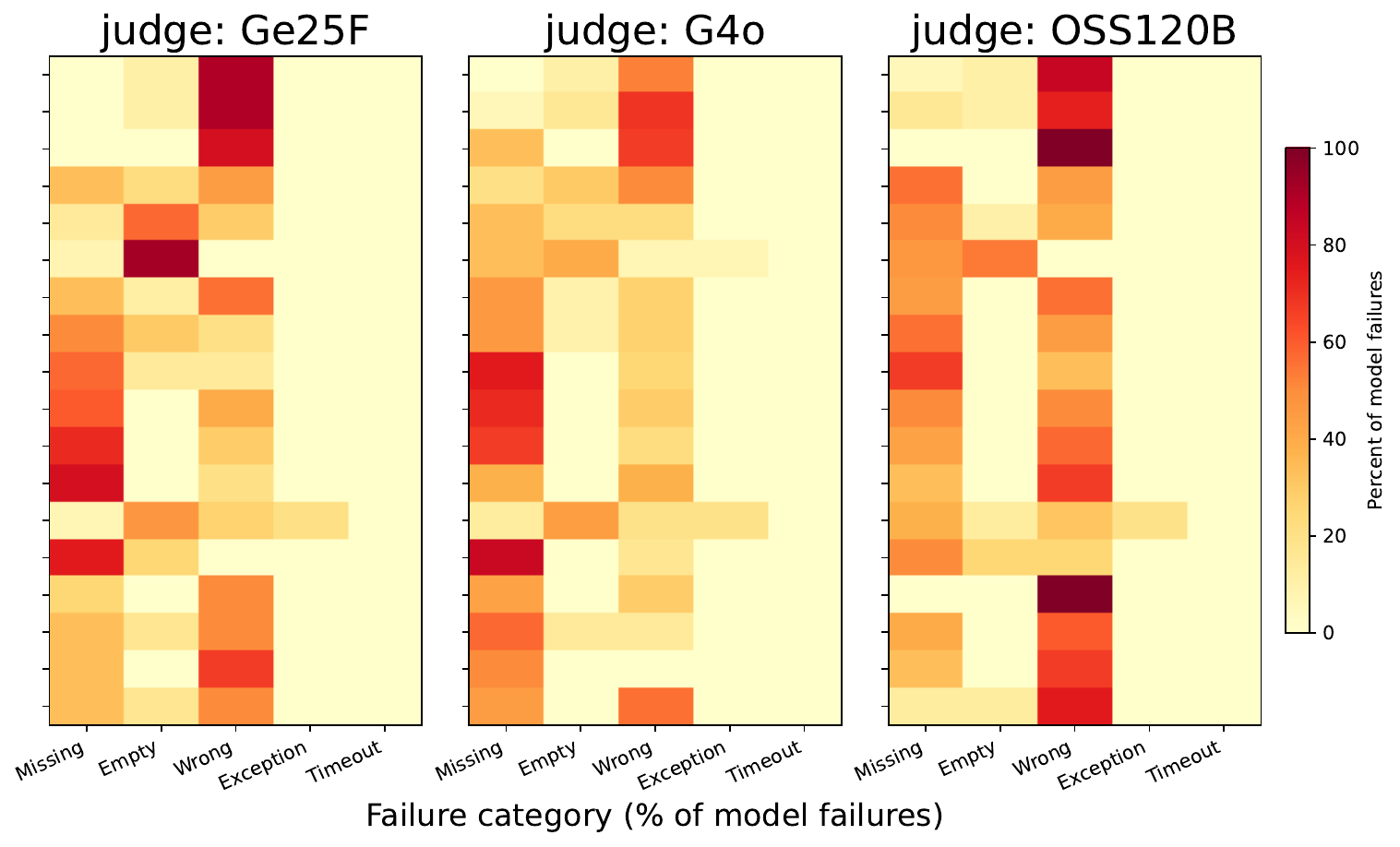}
        \caption{Offline failure attribution via LLM-as-judge (Gemini-2.5-Flash, GPT-4o, GPT-OSS-120B).
        }
        \label{subfig:failure_offline}
    \end{subfigure}
    \caption{
    Failure attribution using different LLM judges.
    We prioritize online attribution when possible, as it incorporates runtime signals unavailable offline, such as execution stalls and incomplete system outputs.
    For offline attribution, judges are provided with the final MAS response, the corresponding gold answer, and an identical failure taxonomy.
    Despite controlling inputs, attribution results exhibit substantial variance across judge models, highlighting the inherent subjectivity and instability of LLM-based failure attribution.
    }
    \label{fig:model-failure}
\end{figure*}

We further break down failure causes by model in \autoref{subfig:failure_gpt4o}, which reveals distinct, model-specific failure signatures. 
Rather than failing uniformly, different LLM backends exhibit characteristic behaviors when errors occur.

\para{Model-specific failure patterns.}

\noindent\begin{itemize}
    \item \textbf{Gemini-2.0-flash-lite} predominantly fails by producing underspecified or incomplete outputs, where a response is returned but lacks sufficient detail to satisfy task requirements.
    \item \textbf{Gemini-2.5-flash-lite} exhibits a more conservative failure mode, frequently abstaining and returning empty or null outputs when uncertain.
    \item \textbf{GPT-4o-mini} tends to produce fully formed but factually incorrect responses, committing confidently to wrong entities or facts rather than omitting answers.
\end{itemize}

These distinct failure signatures indicate that MAS failures are not only model-dependent, but also shaped by how agent architectures interpret and propagate partial outputs. 
Consequently, failures emerge as execution-path–dependent phenomena rather than isolated faults attributable to a single component.

\rqbox{MAS failures predominantly manifest as silent semantic errors, with distinct, model-specific failure signatures that are amplified by execution dynamics.}

\para{Divergent failure attribution across LLM-as-judges.}
To assess the reliability of LLM-as-judge–based failure attribution, we perform offline analysis using three additional judge models, each provided with the final MAS response and the corresponding gold answer. 
As shown in \autoref{subfig:failure_offline}, offline attribution struggles to correctly identify system-level failures, such as exceptions or timeouts, due to the absence of runtime execution signals.

For example, a MAS execution may enter a non-terminating review loop that repeatedly generates responses containing the correct answer but never produces a valid final output. 
During online execution, such behavior is correctly identified as a failure, since the task does not terminate successfully. 
In contrast, an offline judge, which only observes the final response and history, may incorrectly classify the execution as successful because the correct answer appears in the trace.

Even for semantic-based gray failures, where judges often agree on whether an execution is broadly correct or incorrect (e.g., all judges consistently identify CRAG executions as successful), substantial divergence arises in the attribution of failure \emph{types}. 
For instance, when a MAS responds with:  
\emph{“I am sorry, I cannot answer this question. The available tools do not have the functionality to determine the country of a member of the Gujarat Legislative Assembly and parliament.”}  
the \texttt{gpt-oss-120b} judge classifies this outcome as an \emph{empty prediction}, whereas \texttt{gemini-2.5-flash} attributes it to a \emph{wrong fact/entity}.  

These discrepancies highlight that, even under identical inputs and failure definitions, LLM-based judges may disagree on fine-grained failure attribution, underscoring the inherent subjectivity and instability of offline, semantics-only failure analysis.

\subsection{How does tool usage impact cost and accuracy?}
\label{sec:mas-tool-impact}

A common assumption in LLM-based MAS design is that enabling external tools should improve task performance. 
By equipping agents with additional information sources or capabilities, one would expect higher-quality outputs and, consequently, improved accuracy. 
However, our results indicate that the impact of tool usage is highly dependent on the underlying agent architecture.

\begin{figure}
\centering
    \begin{subfigure}[t]{0.86\columnwidth}
    \includegraphics[width=\columnwidth]{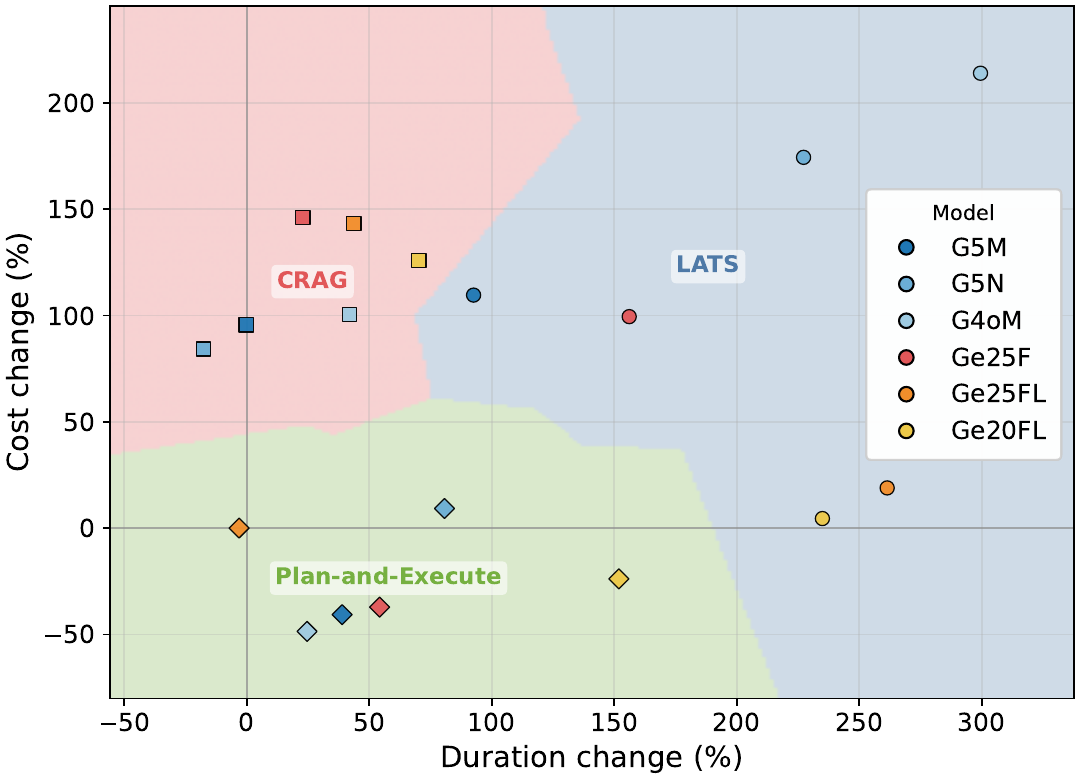}
        \caption{Impact of enabling the Web Search tool. We measure the deltas in execution latency and monetary costs, after enabling the web search tool.
        }
        \label{fig:tavily_cost_diff}
    \end{subfigure}
    \hfill
    \begin{subfigure}[t]{0.66\columnwidth}
        \includegraphics[width=\columnwidth]{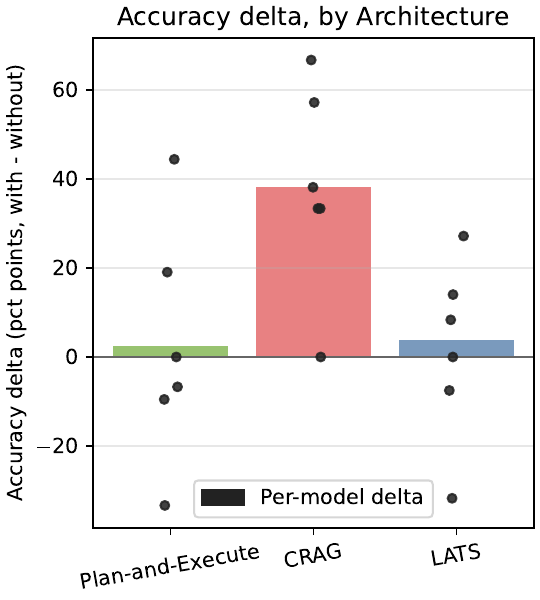}
        \caption{Accuracy deltas after enabling web search.}
        \label{fig:tavily_acc_diff}
    \end{subfigure}
    \caption{
    Resource cost versus accuracy across agent architectures.
    Task-specific designs such as CRAG achieve comparable accuracy with lower resource cost than more general architectures like LATS.
    }
    \label{fig:rq1}
\end{figure}

\para{Enabling web search commonly increases resource consumption.}
Overall, enabling external tools tends to increase resource consumption, but accuracy gains are not uniform across architectures. 
As shown in \autoref{fig:tavily_cost_diff}, tool usage introduces different overheads depending on how tools are integrated into the execution workflow. 
For CRAG, external tools primarily increase monetary cost, with a median cost increase of \$0.0010 per task and a modest median latency increase of 8.1\,s, reflecting additional retrieval and processing steps. In contrast, Plan-and-Execute experiences a substantial increase in task duration, with a median latency increase of 34.1\,s, while its monetary cost slightly decreases, indicating that overhead is shifted toward longer execution rather than additional token usage. 
LATS exhibits the highest overall overhead, with tool usage increasing both execution time and cost, suggesting compounded interaction and coordination overheads.

\para{When web search reduces task duration.}
While external tools typically introduce additional overhead, we observe notable outliers where web search reduces overall execution cost and latency. 
In particular, for CRAG with \texttt{gpt-5-nano}, enabling web search results in faster task completion (by approximately 2\,s on average). This effect arises because, in the absence of external evidence, the model tends to generate longer, more speculative responses, increasing both token usage and per-round LLM latency.
Trace-level analysis confirms this behavior: in no-search executions, the generator and grader produce longer outputs, substantially increasing per-call latency, whereas providing web evidence shortens responses and reduces LLM latency (median generator time 11.2\,s to 6.1\,s). 
As a result, CRAG with web search achieves 13.9\% lower mean task duration despite the additional retrieval step, indicating that external context can reduce speculative reasoning and offset tool overhead.

\para{When web search reduces planning cost.}
For Plan-and-Execute, enabling web search often leads to a net reduction in cost, as external evidence allows the planner to generate more concrete and concise plans. 
Without web search, the planner tends to produce longer, speculative plans, and the replanner emits more verbose messages to justify or revise these plans, inflating token usage.

Trace-level evidence supports this observation: across models, planner messages are substantially shorter when web search is enabled (e.g., average planner tokens drop from over 1{,}500 to a few hundred per call), and replanner turns are also more concise. 
Although the number of planning or replanning iterations may remain similar -- or even increase slightly -- the reduction in per-turn token usage outweighs the cost of the additional web retrieval step, resulting in lower overall execution cost.

\rqbox{By providing external context, tools can reduce speculative generation, lowering inference time and cost.}

\para{Web search boosts accuracy, but non-uniformly across architectures.}
Tool usage yields markedly different outcomes across agent architectures. 
As shown in \autoref{fig:tavily_acc_diff}, CRAG consistently benefits from external tools, achieving a median accuracy improvement of 35.7\% and improving accuracy in 83.3\% of evaluated runs. 
In contrast, Plan-and-Execute loss minor median accuracy, with improvements observed in only one third of runs. 
LATS shows marginal and unstable gains, with a median accuracy improvement of 4.2\% and positive effects in only half of the cases.

In conclusion, these accuracy trends align with the associated cost and latency overheads. 
CRAG incurs only modest increases in cost and execution time, whereas Plan-and-Execute primarily shifts overhead to longer execution latency, and LATS experiences increases in both cost and latency. 
Together, these results indicate that external tools improve MAS performance only when the underlying architecture can incorporate them without amplifying execution complexity or instability.

\rqbox{External tools improve accuracy only when the agent architecture can integrate them without amplifying execution overhead or variance.}

\section{Discussion}

While \sysname already provides valuable insights into the behavior of LLM-based MAS, significant opportunities for extension remain.

\subsection{Limitation}
\para{Generalizability.}
Given the inherent heterogeneity (\textbf{D1}) of LLM-based MAS, it is difficult to identify a single canonical architecture or execution pattern that generalizes across all agentic systems. 
The design space of MAS continues to evolve rapidly, with new coordination strategies, tooling abstractions, and execution models emerging at a fast pace.

While \sysname is designed to cover a diverse set of widely used MAS architectures and workflows, the insights derived from our evaluation are necessarily grounded in the specific instances and configurations studied. 
As a result, some findings may not directly transfer to future MAS designs or to application domains not represented in our benchmark. 
In particular, advances in agent orchestration or model capabilities may invalidate certain observations over time, highlighting the need for continuously evolving benchmarks alongside the MAS ecosystem.

\para{Overhead of telemetry.}
\sysname incorporates fine-grained telemetry to examine per-step behavior in LLM-based MAS and derive detailed insights into execution dynamics. 
However, such instrumentation introduces profiling overhead that may degrade system performance. 
Before real-world deployment, \sysname must therefore optimize telemetry collection to minimize overhead, for example, through sampling strategies, adaptive logging, or lightweight monitoring mechanisms.

\subsection{Future works}
\para{Automated integration for MAS instances.}
Currently, \sysname includes a limited representative set of MAS examples, which may not fully capture the diversity and complexity of real-world deployments.
To improve representativeness, the benchmark must incorporate a broader range of MAS examples that reflect the diversity of real-world deployments and use cases. 
At present, fine-grained telemetry is enabled through ad-hoc instrumentation tailored to individual MAS frameworks. 
As future work, we plan to develop an automated translation layer that maps heterogeneous agent implementations into a uniform execution representation, enabling systematic behavior capture with minimal manual intervention. 
Such automated integration would also lower the barrier for external contributions, allowing developers to more easily evaluate their own MAS implementations using our test suite.

\para{Monolith vs. distributed.} Similar to the relationship between monolithic applications and microservices in traditional software architectures, LLM-based MAS can be deployed either as a single, unified system or as a collection of distributed agents communicating over a network. Distributed deployments could bring benefits such as improved fault tolerance, scalability, and modularity. However, they also introduce challenges related to network latency, synchronization, and consistency. Future work could explore the trade-offs between monolithic and distributed MAS architectures, evaluating their performance, reliability, and resource utilization under various workloads and deployment scenarios. Also, it would be interesting to investigate the impact of the underlying network infrastructure on the behavior and performance of distributed LLM-based MAS.

\para{MAS-specific failure attribution.}
The inherent non-determinism of LLMs introduces failure modes that are rarely encountered in traditional deterministic systems. 
When multiple agents are composed into a pipeline, these effects are further amplified, increasing the likelihood of inconsistent or emergent failure behaviors.
Such phenomena are already observed in our evaluation.
For instance, in the Plan-and-Execute architecture, we identify recurring execution patterns in which the executor successfully retrieves and returns the gold answer, yet the replanner repeatedly rejects the intermediate result. 
This mismatch prevents the system from reaching a terminal state, ultimately leading to timeouts despite the presence of a correct solution in the execution trace.
These observations highlight the difficulty of failure attribution in LLM-based MAS. 
Due to the extensive fault space induced by LLM heterogeneity (\textbf{D1}), which grows combinatorially as multiple agents and models interact, failures often cannot be localized to a single component or decision point. 
Developing principled failure taxonomies and robust attribution mechanisms for such systems therefore remains an important direction for future work.

\para{Communication mechanisms.}
Our experiments reveal substantial variation in how different MAS frameworks implement inter-agent communication. 
In many of the frameworks we evaluate, agents primarily interact through structured function calls. 
Others rely on a shared global scratchpad that allows agents to read from and write to a common intermediate state.

For interactions beyond a single host or for accessing external data sources, some frameworks additionally support standardized communication protocols, such as agent-to-agent (A2A)~\cite{a2aProtocol} messaging or the Model Context Protocol (MCP)~\cite{mcp}. 
These differences in communication mechanisms introduce distinct execution semantics and coordination patterns, yet their impact on system performance, robustness, and failure behavior remains largely unexplored. 
This observation highlights an open research area in understanding how communication design choices influence the behavior of LLM-based MAS.

\para{Parallelism and coordination effects.}
Parallelism fundamentally alters the execution dynamics of LLM-based MAS, affecting not only throughput and resource utilization but also coordination behavior and failure modes. 
While parallel execution and load balancing are well-established techniques in traditional systems, their impact in asynchronous MAS remains poorly understood.

Existing work has extensively studied single-LLM optimizations, such as speculative inference and parallel decoding~\cite{leviathan2023fast, chen2023accelerating}, to improve accuracy or cost–performance trade-offs. 
However, it is unclear how these techniques translate to multi-agent settings, where multiple agents may operate concurrently and interact through shared state or tools.

In particular, the effects of parallel agents with overlapping or partially redundant roles are not yet well characterized. 
Such configurations may introduce new coordination overheads, contention, or emergent behaviors that differ fundamentally from single-model parallelism, highlighting an important direction for future investigation.

\para{Framework overhead investigation.} In our evaluation, we observe that kagent~\cite{kagent}, a framework designed to facilitate building distributed LLM-based multi-agent systems, can incur non-trivial communication overhead and may also trigger operational failures (e.g., a disk-full error on the kagent controller node). Future work should systematically characterize the overheads introduced by MAS frameworks and quantify their impact on end-to-end performance and reliability.

\section{Conclusion}
We argue that LLM-based multi-agent systems (MAS) must be evaluated not merely by task completion, but as complex systems characterized by dynamic, stochastic execution. To this end, we introduce \sysname, an open-source evaluation suite that standardizes the configuration and execution of heterogeneous MAS while exporting fine-grained, system-level telemetry to enable cross-stack comparison.

Our evaluation of 12 representative MAS instances reveals that while agentic workflows are often \emph{structurally} stable, they exhibit significant \emph{temporal} instability, driving high run-to-run variance in latency, cost, and failure modes. Crucially, we find that MAS architecture dominates backend model and toolset choices in determining resource profiles, reproducibility, and the cost--latency--accuracy trade-off. These findings indicate that optimizing reliability and efficiency in agentic systems is fundamentally an architectural challenge, necessitating benchmarks that prioritize deep execution visibility over simple application-level scores.

Looking ahead, we plan to extend \sysname to support distributed architectures and automated agent integration, while refining failure attribution to better diagnose stochastic errors. Our ultimate goal is to establish standardized observability contracts, ensuring that benchmarking keeps pace with the evolving complexity of agentic systems.\newpage
\bibliographystyle{ACM-Reference-Format}
\bibliography{bibs}
\clearpage

\appendix

\section{Appendix}

\subsection{Details of post-processing component}
\label{subsec: Details of Post Processing Component}

We build an observation component that characterizes MAS behavior across benchmark tasks and configurations. By default, we generate a common set of plots for each workload. We report results both for individual executions and aggregated across multiple runs. Because the full set of figures is large, we plan to release them as a dataset and provide only a brief summary here.

\begin{itemize}
    \item \textbf{Token consumption}. For the single run, we plot the token consumption of each agent over time, including prompt tokens and completion tokens; we also plot the total token consumption of all agents over time. For multiple runs, we plot the average total token consumption of all agents over time.
    \item \textbf{Delay}. For the single run, we plot the end-to-end delay of the whole system, and decompose it into agent processing delay, agent-to-LLM communication delay, and agent-to-agent communication delay. For multiple runs, we plot the average end-to-end delay of the whole system. We plot the breakdown of the average delay into different components. We also plot a flame graph for delay.
    \item \textbf{CPU and memory usage}. For the single run, we plot the time series of CPU and memory usage of the whole system. We also give a correlation analysis between CPU/memory usage and system events (e.g., agent invocations, LLM calls, etc.). For multiple runs, we plot the mean, peak, and minimum CPU and memory usage of the whole system.
    \item \textbf{Message size}. For the single run, we plot the average input and output message size of each agent for both agent-to-agent messages and agent-to-LLM messages. For multiple runs, we plot the total input and output message size of agent-to-agent messages and agent-to-LLM messages. We also plot per-agent, per-agent-pair, and per-agent-to-LLM message sizes.
    \item \textbf{Call graph}. We visualize the call graph of the agents in the system. We also show the similarity between the call graphs of different runs using graph similarity metrics. We use two similarity metrics: Jaccard similarity and Largest Common Subgraph (LCS) similarity. Jaccard similarity measures the similarity between two sets of edges in the call graphs, while LCS similarity measures the sequence similarity of the edges in the call graphs.
\end{itemize}

\subsection{Collector implementation.}

\subsubsection{Telemetry field collection}
\label{subsec: otel template}

The following listing (Listing~\ref{lst:json-trace}) shows a collection of telemetry fields in \sysname.

\begin{lstlisting}[language=json, caption={Telemetry field collection}, label={lst:json-trace}]
[
  {
    "trace_id": "<32-hex-trace-id>",
    "span_id": "<16-hex-span-id>",
    "parent_span_id": "<16-hex-parent-span-id-or-null>",
    "name": "<operation-name>",
    "agent_name": "<agent-name>",
    "start_time": 0,
    "end_time": 0,
    "duration_ns": 0,
    "kind": "<INTERNAL|SERVER|CLIENT|PRODUCER|CONSUMER>",
    "status": {
      "status_code": "<UNSET|OK|ERROR>",
      "description": "<optional-description>"
    },
    "attributes": {
      "gen_ai.operation.name": "<call_llm|execute_tool|invoke_agent>",
      "gen_ai.system": "<provider>",
      "gen_ai.agent.name": "<agent-name>",
      "gen_ai.agent.description": "<optional-description>",
      "gen_ai.request.model": "<model>",
      "gen_ai.conversation.id": "<conversation-id>",
      "gen_ai.tool.name": "<tool-name>",
      "gen_ai.tool.type": "<FunctionTool|Builtin>",
      "gen_ai.tool.call.id": "<tool-call-id>",
      "gen_ai.tool.description": "<tool-description>",
      "gen_ai.usage.input_tokens": 0,
      "gen_ai.usage.output_tokens": 0,
      "gen_ai.usage.total_tokens": 0,
      "gen_ai.llm.call.count": 0,
      "gen_ai.mcp.call.count": 0,
      "gen_ai.response.finish_reasons": [],
      "mcp.server": "<server-name>",
      "mcp.tool": "<tool-name>",
      "gcp.vertex.agent.llm_request": "<raw-request-json>",
      "gcp.vertex.agent.llm_response": "<raw-response-json>",
      "gcp.vertex.agent.tool_call_args": "<tool-call-args>",
      "gcp.vertex.agent.tool_response": "<tool-response>",
      "gcp.vertex.agent.invocation_id": "<invocation-id>",
      "gcp.vertex.agent.session_id": "<session-id>",
      "gcp.vertex.agent.event_id": "<event-id>",
      "agent.log": "<optional-log-line>",
      "agent.retry.attempt_number": 0,
      "agent.retry.trigger": "<quality|relevance_guard|guard_fail|timeout|system|upstream>",
      "agent.retry.previous_span_id": "<16-hex-span-id-or-null>",
      "agent.retry.reason": "<optional-retry-trigger>",
      "run.outcome": "<success|failure>",
      "run.outcome_reason": "<optional-reason>",
      "run.judgement": "<correct|wrong|unknown>",
      "run.judgement_reason": "<optional-reason>",
      "agent.failure.category": "<guard|quality|system|timeout|upstream>",
      "agent.failure.reason": "<free-text>",
      "agent.output.useless": false,
      "agent.output.useless_reason": "<free-text>",
      "communication.input_message_size_bytes": 0,
      "communication.output_message_size_bytes": 0,
      "communication.total_message_size_bytes": 0
    },
    "communication": {
      "is_in_process_call": false,
      "input_message_size_bytes": 0,
      "output_message_size_bytes": 0,
      "total_message_size_bytes": 0
    },
    "resource": {
      "attributes": {
        "service.name": "<service-name>",
        "service.version": "<semver>",
        "deployment.environment": "<local|dev|staging|prod>",
        "telemetry.sdk.name": "<sdk-name>",
        "telemetry.sdk.language": "<language>",
        "telemetry.sdk.version": "<version>",
        "host.name": "<optional-host>"
      }
    }
  }
]
\end{lstlisting}

\subsubsection{Lacking of standardized observability contracts.}
Even when a common observability schema (e.g., OTEL) is imposed, orchestration stacks differ substantially in which telemetry signals are surfaced, transformed, or suppressed.
While some frameworks propagate execution metadata such as token usage, termination reasons, or payload sizes to application-level hooks, others consume these signals within internal execution layers without exposing them externally. 
As a result, identical agent workflows may exhibit markedly different observability characteristics depending on the combination of model backend, transport mechanism, and orchestration framework.

A key source of this discrepancy is that, unlike generated text, token usage is not treated as a first-class execution artifact with a well-defined exposure contract, but rather as auxiliary metadata. 
Consequently, whether token usage is observable depends jointly on (i) the underlying model API and its response schema, (ii) the transport layer through which inference results are delivered (e.g., streaming versus non-streaming), and (iii) the framework’s instrumentation and log-propagation strategy. 
In the absence of an agreed-upon contract, each layer independently decides how token usage is represented and whether it is forwarded, making end-to-end observability fragile and stack-dependent.

\para{Backend- and modality-dependent loss of usage metadata.}
This lack of standardization manifests across both model backends and invocation modalities. 
For example, Gemini and Vertex AI do expose token usage information, but under response layouts and terminology that differ from OpenAI- or Anthropic-style APIs. 
Token usage may be reported through backend-specific metadata fields (e.g., reporting generation-side token usage as \emph{candidate tokens}, rather than OpenAI-style output or completion tokens~\cite{langchain_vertexai_chatvertexai}), requiring backend-aware parsing logic to recover usage information.
Beyond generation APIs, we further observe that usage metadata may be dropped entirely at the framework level for non-generative calls. 
In LangGraph, embedding model invocations (e.g., \texttt{OpenAIEmbeddings} and \texttt{VertexAIEmbeddings}) do not propagate token usage information, even when the underlying provider APIs support usage accounting. 
In such cases, the framework consumes partial response metadata internally without forwarding it to application-level telemetry or accounting hooks.
As a consequence, orchestration frameworks such as LangGraph and MCP-Agent -- many of which implicitly assume a synchronous, OpenAI-style usage schema -- may fail to capture token usage across a range of execution paths unless explicit, backend- and modality-aware instrumentation is implemented. 
Importantly, these limitations do not arise from agent logic or missing backend signals, but from the absence of a stable, cross-provider observability contract that defines how usage metadata should be structured, preserved, and forwarded across abstraction boundaries.

\para{Implications for MAS benchmarking.}
This inconsistency introduces blind spots in cost and efficiency analysis, particularly in heterogeneous multi-agent settings where different LLM backends coexist. 
It further demonstrates that observability properties cannot be assumed to be model-agnostic, motivating the need for standardized observability contracts that explicitly define which execution signals must be exposed by LLM APIs and agent frameworks.

\end{document}